\documentclass[reprint]{revtex4-1}

\pdfoutput=1 

\usepackage{graphicx} 
\usepackage{amsfonts} 
\usepackage{amssymb} 
\usepackage{amsmath}
\usepackage{color}
\usepackage{amsmath,amssymb,latexsym,epsfig,graphics,epsf,bm}
\usepackage{graphicx}
\usepackage{float}
\usepackage{placeins}
\usepackage[normalem]{ulem}

\usepackage{rotating}
\usepackage{longtable}
\usepackage{multirow}

\newcommand{\br}{\mathbf r}

\newcommand{\bR}{\mathbf R}

\newcommand{\tbR}{\tilde{\mathbf R}}

\newcommand{\be}{\begin{equation}}
\newcommand{\ee}{\end{equation}}

\newcommand{\tU}{\tilde{\Phi}}

\renewcommand{\vec}[1]{{\bf #1}}

\newcommand{\eps}{\varepsilon}

\begin{document}

\title{Hidden scale invariance of metals}
\author{Felix Hummel$^{1}$} 
\author{Georg Kresse$^{1}$} 
\author{Jeppe C. Dyre$^{2}$} 
\author{Ulf R. Pedersen$^{2,3}$} 
\email{ulf@urp.dk} 
\affiliation{$^1$Faculty of Physics and Center for Computational Materials Science, University of Vienna, Sensengasse 8/12, A-1090 Vienna, Austria} 
\affiliation{$^2$DNRF Center "Glass and Time", IMFUFA, Dept. of Sciences, Roskilde University, P. O. Box 260, DK-4000 Roskilde, Denmark} 
\affiliation{$^3$Institute of Theoretical Physics, Vienna University of Technology, Wiedner Hauptstra{\ss}e 8-10, A-1040 Vienna, Austria} 

\date{\today} 
\pacs{...} 
\begin{abstract} 
Density functional theory (DFT) calculations of 58 liquid elements at their triple point show that most metals exhibit near proportionality between thermal fluctuations between virial and potential-energy in the isochoric ensemble. This demonstrates a general ``hidden'' scale invariance of metals making the dense part of the thermodynamic phase diagram effectively one dimensional with respect to structure and dynamics. DFT computed density scaling exponents, related to the Gr{\"u}neisen parameter, are in good agreement with experimental values for 16 elements where reliable data were available. Hidden scale invariance is demonstrated in detail for magnesium by showing invariance of structure and dynamics. 
Computed melting curves of period three metals follow curves with invariance (isomorphs). 
The experimental structure factor of magnesium is predicted by assuming scale invariant inverse power-law (IPL) pair interactions. However, crystal packings of several transition metals (V, Cr, Mn, Fe, Nb, Mo, Ta, W and Hg), most post-transition metals (Ga, In, Sn, and Tl) and the metalloids Si and Ge cannot be explained by the IPL assumption. Thus, hidden scale invariance can be present even when the IPL-approximation is inadequate. The virial-energy correlation coefficient of iron and phosphorous is shown to increase at elevated pressures. Finally, we discuss how scale invariance explains the Gr{\"u}neisen equation of state and a number of well-known empirical melting and freezing rules.
\end{abstract}

\maketitle 

\section{Introduction}

Scale invariance plays an important role in many branches of science. It greatly simplifies a given phenomenon by reducing the parameter space and introducing universalities over length or time scales. Some examples of this are the size distribution of earth quakes \cite{bak2002}, Brownian motions of microscopic particles \cite{mandelbrot1968}, cosmic microwave background radiation \cite{guth1982,hawking1982}, and biological fractal structures \cite{mandelbrot83} such as those of lung tissue or Romanesco broccoli. In the physics of matter, scale invariance controls the properties of a fluid near the gas-liquid critical point \cite{pelissetto2002}. This paper establishes an approximate ``hidden'' scale invariance in the dense liquid part of the thermodynamic phase diagram of certain elements from {\it ab initio} computations. While the above classic examples of scale invariance cover several decades of variations, the scale invariance of condensed matter covers much smaller length-scale differences. Nonetheless, hidden scale invariance implies that the phase diagram becomes effectively one-dimensional. 

Specifically, we have performed {\it ab initio} density functional theory (DFT) computations on 58 liquid elements at their triple point. We infer hidden scale invariance of metals from strong correlations in thermal fluctuations of virial $W$ (the potential part of the pressure) and internal potential energy $U$. We find that metals in general posses hidden scale invariance.
These results give the first {\it ab initio} quantum-mechanical confirmation of the picture proposed in Refs. \onlinecite{ped08,I,II,III,IV,ing12,dyr14}, according to which some systems have strong virial potential-energy correlations in the condensed phases, whereas this is not the case for other systems -- typically those with strong directional bonding. Liquids belonging to the former class of systems were originally referred to as ``strongly correlating'', but to avoid confusion with strongly correlated quantum systems the term Roskilde (R) simple system is now sometimes used \cite{mal13,pra14,fle14,hen14,pie14,hey15}. Although the focus here is on monatomic systems, experimental and molecular-dynamics simulation results have shown that (R) simple systems include some molecular van der Waals bonded systems \cite{gun11}, polymers \cite{rol05,rol10}, and crystalline solids \cite{alb14}. Returning to the elements, molecular liquids like $N_2$ \cite{II} and the noble gases \cite{ped08} are also expected to be simple \cite{I}.

%


What is the possible microscopic origin of hidden scale invariance?
To answer this, we start by noting that at high pressure the dominant interatomic forces are harshly repulsive \cite{waals1873,bor32,bor39,wid67,wca,gub71,bar76,and76,wri81}.  These forces can be modeled approximately by scale-invariant inverse-power-law (IPL) pair-potentials terms $\propto r^{-n}$ \cite{hoo71,hoo72,hiw74,sti75,ros76,you77,ros83,you91,pre05,bra06,hey07,hey08,ped10,bra11,khr11a,tra14} in which $r$ is the interparticle distance, plus a density-dependent constant $g(\rho)$ taking long-ranged attractive interactions into account \cite{waals1873,wid67}. By Euler's theorem for homogeneous functions, there is exact correlation between the fluctuations of virial and potential-energy in the IPL case. 
%
Strong correlations between virial and potential energy are, however, not necessarily a consequence of approximately scale-invariant pair interactions, but rather a property of the multidimensional energy-function $U(\bR,V)$ {\it per se}. This fact motivates the present paper in which we show from first-principles computations that many metals obey Eq.\ (\ref{WUFluc}) also at {\it low pressure}, i.e., when virtually uncompressed compared to the zero-pressure state. 

For metals, the outer electrons result in complex many-body interactions, which this study takes explicitly into account. At low pressure hidden scale invariance is nontrivial since the forces on the atoms are not at all dominated by pairwise additive repulsive IPL-type forces \cite{hafner86}. Hidden scale invariance have been found to be absent if the multidimensional energy function $U(\bR,V)$ have prominent contributions characterized by two or more length scales \cite{I,II}. Examples of this are water with hydrogen bonds and core repulsions, and the 
Dzugutov model with core repulsions supplemented by repulsive forces on second-nearest neighbors \cite{I,II}. However, it is possible to have multi-body interactions and hidden scale 
invariance.
Below we do not make any assumptions on the potential-energy landscape and verify hidden scale invariance for the majority of metals directly from a quantum-mechanical description. Density Functional Theory (DFT) currently provides the best means to do so \cite{koh99,pop99,coh12,bur12}, and although DFT can only approximate the true electronic behavior, it provides accurate predictions on a broad scale, such as crystal structures, densities and melting points.


The remainder of the paper is organized as follows: Sec.\ \ref{theory} gives the theoretical background of hidden scale invariance and Sec.\ \ref{method} gives details on the {\it ab initio} DFT method. 
Results are presented in Sec. \ref{results}, and 
in Sec. \ref{discussion} we relate our findings to Gr{\"u}neisen equation of state, and well-known empirical melting and freezing rules.

\section{Theory: Hidden scale invariance of condensed matter}\label{theory}

Certain condensed-matter systems are characterized by a ``hidden scale invariance'' as reflected in the following approximate representation of the potential-energy function at a density $\rho=N/V$ \cite{IV,ing12,dyr14}: 
\be\label{hsi}
U(\bR)\cong h(\rho)\tU(\tbR)+Ng(\rho)\,,
\ee
where the coordinates of the system's $N$ particles have been merged into a single vector, $\bR\equiv (\br_1,\br_2,...,\br_N)$, and the reduced coordinate is $\tbR\equiv\rho^{1/3}\bR$.
The intensive functions $h(\rho)$ and $g(\rho)$ both have dimension energy and $\tU$ is a dimensionless, state-point-independent function of the dimensionless variable $\tbR$, i.e., a function that involves no lengths or energies. Physically, Eq.\ (\ref{hsi}) means that a change of density to a good approximation leads to a linear affine transformation of the high-dimensional potential-energy surface. Thus if temperature is adjusted in proportion to $h(\rho)$, state points in the thermodynamic phase diagram are arrived at for which the molecules according to Newton's laws move in the same way, except for a uniform scaling of space and time. Sets of such state points are referred to as isomorphs and along the isomorphs structure and dynamics are identical in properly reduced units to a good approximation \cite{III}. The local slope $d \ln(T)/d\ln(\rho)$ of the isomorph is given by 
\be
\gamma(\rho)=d\ln(h(\rho))/d\ln(\rho),
\ee
and is refered to as the density scaling exponent \cite{alb02,rol05,alb06,IV,rol10,flo11,ing12,dyr14,mau14}. Thus, the phase diagram becomes effectively one-dimensional, and density and temperature merge into a single parameter.
%
%

Hidden scale invariance is revealed in thermal fluctuations at a single state point:
There are two contributions to the pressure, the ideal-gas pressure $p_{\rm id}$ -- a term that is always present and which only depends on the particles' (atoms') velocities -- plus a term deriving from the interaction between the particles. The latter is the so-called virial $W$, and it only depends on the particles' coordinates. $W$ is an extensive quantity of dimension energy and the general pressure relation is $p=p_{\rm id}+W/V$.
Hidden scale invariance dictates that fluctuations of virial and potential energy are strongly correlated in the $NVT$ ensemble \cite{ped08,I,II,IV}:
\begin{equation}\label{WUFluc}
W(\bR,V) \cong \gamma(\rho)U(\bR,V)+\textrm{const.}
\end{equation}
where the microscopic virial is defined by $W(\bR,V)\equiv -V \partial{U(\bR,V)}/ { \partial V}$ \cite{han13,tildesley}.
Below we use the Person correlation coefficient
\be \label{person}
 R = \langle \Delta W(\bR) \Delta U(\bR) \rangle/\sqrt{\langle [\Delta W(\bR)]^2 \rangle\langle [\Delta U(\bR)]^2 \rangle}
\ee 
of virial and potential energy fluctuations to determine to what degree the multi body energy function $U(\bR)$ have hidden scale invariance. Here, $\langle\dots\rangle$ indicate a thermodynamic average in the NVT ensemble and $\Delta$'s are differences to average values. $R$ is an intensive quantity with a value between -1 and 1. A value close to 1 indicate hidden scale invariance.
The density scaling exponent is given by \cite{III}
\be \label{gammaWU}
 \gamma = \langle \Delta W(\bR) \Delta U(\bR) \rangle/\langle [\Delta U(\bR)]^2 \rangle.
\ee

We have previously conjectured that metals possess a hidden scale invariance, but this was based on assuming Lennard-Jones type pair potentials \cite{ped08}.
Ref. \cite{I} showed via the Effective Medium Approximation that pure copper, as well as a magnesium alloy, exhibit hidden scale invariance.
We do not make such assumptions in this paper, as detailed in the following section.

\section{Method: Density Functional Theory}\label{method}
DFT molecular-dynamics is a computationally efficient method with a quantum-mechanical treatment of the electron-density field. Examples of recent successes of DFT are computation of anharmonic contribution due to phonon-phonon interactions for fcc crystals of metals \cite{PhysRevLett.114.195901}, dynamics of water dissociative chemisorption on a Ni surface \cite{PhysRevLett.114.166101}, and accurate computations of the melting points of period-three metals \cite{PhysRevB.88.094101}. 

In this paper we present calculations of 58 elements using the Vienna Ab-initio Simulation Package (VASP) \cite{PhysRevB.54.11169} employing the projector augmented wave (PAW) method \cite{PhysRevB.50.17953} with the frozen-core assumption and the Perdew-Burke-Enzerhof (PBE) exchange-correlation functional \cite{PhysRevLett.100.136406}. Dispersion corrections are not included, but these are not expected to be important for metals near the melting point \cite{PhysRevB.88.094101}. 
We use periodic simulation cells containing 125 atoms for most elements except
for the elements Li, Na, Mg and Al where we use 256, 250, 384 and 256 atoms, respectively.
Initial equilibration trajectories cover between 9 to 24\,ps corresponding to several structural relaxation times.
Constant temperature is obtained with the Langevin thermostat with a coupling
time of 1\,ps.
Table \ref{tab:details} contains the electronic
configurations of the calculated electrons of the employed potentials as well
as the cutoff energy of the plane wave basis set. All calculations were
done non-spin polarized, and the Brillouin zone was
sampled at its center  the $\Gamma$ point.
Statistical uncertainties are estimated by dividing MD trajectories into statistically independent blocks.
Details on estimating DFT triple points are given in the Appendix.

\begingroup
\squeezetable
\begin{table*}[p]
\begin{tabular}{|rl|r|l|r|rl|rl|r|r|r|r|r|}
  \hline
  \multicolumn{2}{|c|}{\multirow{2}{*}{Element}} &
    \multicolumn{1}{c|}{\multirow{2}{*}{$N$}} &
    \multicolumn{1}{c|}{\multirow{2}{*}{Electrons}} &
    \multicolumn{1}{|c|}{$E_{\rm max}$} &
    \multicolumn{1}{|c}{$T$} &
    \multicolumn{1}{c|}{[exp.]} &
    \multicolumn{1}{c}{$\rho$} &
    \multicolumn{1}{c|}{[exp.]} &
    \multicolumn{1}{c|}{$p$} &
    \multicolumn{1}{c|}{\multirow{2}{*}{$R$}} &
    \multicolumn{1}{c|}{\multirow{2}{*}{$\gamma$}} &
    \multicolumn{1}{c|}{\multirow{2}{*}{$\sigma$}} &
    \multicolumn{1}{c|}{\multirow{2}{*}{$R_{\rm IPL}$}} \\
  \multicolumn{2}{|c|}{} &
     &
     &
    \multicolumn{1}{|c|}{[eV]} &
    \multicolumn{2}{|c|}{[K]} &
    \multicolumn{2}{c|}{[g/cm$^3$]} &
    \multicolumn{1}{c|}{[GPa]} & & & &  \\
\hline
3 & Li & 256 & 2s$^1$ & 140.0 & 470 & [454] & 0.534 & [0.512] & 0.29 & 0.69(0.08) & 1.0(0.1) & 1.92(0.28) & 0.71 \\
4 & Be & 125 & 2s$^2$ & 308.8 & 1850 & [1560] & 1.570 & [1.690] & -0.27 & 0.75(0.03) & 1.8(0.3) & 1.16(0.09) & 0.50 \\
5 & B & 125 & 2s$^2$2p$^1$ & 318.6 & 2400 & [2349] & 2.245 & [2.08] & -1.69 & 0.10(0.12) & \multicolumn{1}{c|}{---} & \multicolumn{1}{c|}{---} & \multicolumn{1}{c|}{---} \\
6 & C & 125 & 2s$^2$2p$^2$ & 273.9 & 6000 & [4800] & 1.514 & [1.37] & -1.04 & -0.04(0.18) & \multicolumn{1}{c|}{---} & \multicolumn{1}{c|}{---} & \multicolumn{1}{c|}{---} \\
\hline
11 & Na & 250 & 3s$^1$ & 62.1 & 370 & [371] & 0.941 & [0.927] & -0.02 & 0.83(0.02) & 1.9(0.1) & 1.25(0.03) & 0.89 \\
12 & Mg & 384 & 3s$^2$ & 98.5 & 900 & [923] & 1.581 & [1.584] & 0.07 & 0.90(0.01) & 2.5(0.2) & 1.15(0.03) & 0.81 \\
13 & Al & 256 & 3s$^2$3p$^1$ & 116.4 & 1000 & [933] & 2.352 & [2.375] & -0.27 & 0.86(0.06) & 4.0(0.4) & 1.02(0.02) & 0.90 \\
14 & Si & 125 & 3s$^2$3p$^2$ & 245.3 & 1700 & [1687] & 2.827 & [2.57] & 0.51 & 0.79(0.03) & 4.1(0.2) & 0.94(0.02) & 0.79 \\
15 & P & 125 & 3s$^2$3p$^3$ & 255.0 & 650 & [317] & 1.850 & [1.74] & -0.43 & 0.01(0.17) & \multicolumn{1}{c|}{---} & \multicolumn{1}{c|}{---} & \multicolumn{1}{c|}{---} \\
16 & S & 125 & 3s$^2$3p$^4$ & 258.7 & 550 & [388] & 1.784 & [1.819] & 0.07 & 0.14(0.09) & \multicolumn{1}{c|}{---} & \multicolumn{1}{c|}{---} & \multicolumn{1}{c|}{---} \\
\hline
19 & K & 125 & 3p$^6$4s$^1$ & 116.7 & 350 & [337] & 0.815 & [0.828] & 0.07 & 0.84(0.12) & 1.6(0.3) & 1.45(0.27) & 0.84 \\
20 & Ca & 125 & 3p$^6$4s$^2$ & 266.6 & 1200 & [1115] & 1.378 & [1.378] & -0.44 & 0.80(0.06) & 1.9(0.1) & 1.20(0.03) & 0.83 \\
21 & Sc & 125 & 4s$^2$3d$^1$ & 154.8 & 1900 & [1814] & 2.800 & [2.80] & -0.38 & 0.63(0.24) & 1.4(0.5) & 1.43(0.56) & 0.81 \\
22 & Ti & 125 & 4s$^2$3d$^2$ & 178.3 & 3900 & [1941] & 4.161 & [4.11] & -0.48 & 0.78(0.08) & 2.0(0.2) & 0.96(0.04) & 0.88 \\
23 & V & 125 & 3p$^6$4s$^2$d$^3$ & 263.7 & 2500 & [2183] & 5.773 & [5.5] & -0.13 & 0.81(0.07) & 2.6(0.6) & 1.02(0.12) & 0.81 \\
24 & Cr & 125 & 3p$^6$4s$^1$d$^5$ & 265.7 & 2600 & [2180] & 6.735 & [6.3] & -1.55 & 0.90(0.04) & 3.3(0.9) & 0.98(0.07) & 0.83 \\
25 & Mn & 125 & 4s$^2$3d$^5$ & 269.9 & 2400 & [1519] & 7.973 & [5.95] & 0.66 & 0.93(0.02) & 3.6(0.2) & 0.95(0.03) & 0.72 \\
26 & Fe & 125 & 3p$^6$4s$^2$3d$^6$ & 267.9 & 2400 & [1811] & 8.338 & [6.98] & 0.03 & 0.95(0.02) & 3.6(0.1) & 1.00(0.01) & 0.90 \\
27 & Co & 125 & 4s$^2$3d$^7$ & 268.0 & 1870 & [1768] & 8.924 & [7.75] & 0.22 & 0.93(0.01) & 3.5(0.1) & 1.06(0.01) & 0.94 \\
28 & Ni & 125 & 4s$^2$3d$^8$ & 269.5 & 2000 & [1728] & 8.189 & [7.81] & -0.17 & 0.92(0.03) & 3.5(0.3) & 1.03(0.02) & 0.96 \\
29 & Cu & 125 & 4s$^1$3d$^{10}$ & 295.4 & 1480 & [1358] & 8.020 & [8.02] & -0.72 & 0.90(0.02) & 4.1(0.2) & 1.01(0.01) & 0.94 \\
30 & Zn & 125 & 4s$^2$3d$^{10}$ & 276.7 & 760 & [693] & 6.570 & [6.57] & -0.61 & 0.53(0.12) & 3.3(0.9) & 1.03(0.04) & 0.43 \\
31 & Ga & 125 & 4s$^2$4p$^1$ & 134.7 & 500 & [303] & 5.967 & [6.095] & -0.41 & 0.74(0.09) & 3.3(0.5) & 1.09(0.04) & 0.65 \\
32 & Ge & 125 & 4s$^2$3d$^{10}$4p$^2$ & 310.3 & 1250 & [1211] & 5.600 & [5.60] & -1.46 & 0.82(0.15) & 4.8(1.1) & 0.91(0.02) & 0.80 \\
33 & As & 125 & 4s$^2$4p$^3$ & 208.7 & 1300 & [1090] & 5.220 & [5.22] & 0.58 & -0.08(0.13) & \multicolumn{1}{c|}{---} & \multicolumn{1}{c|}{---} & \multicolumn{1}{c|}{---} \\
34 & Se & 125 & 4s$^2$4p$^4$ & 211.6 & 550 & [494] & 3.838 & [3.99] & 0.16 & -0.02(0.21) & \multicolumn{1}{c|}{---} & \multicolumn{1}{c|}{---} & \multicolumn{1}{c|}{---} \\
\hline
37 & Rb & 125 & 4p$^6$5s$^1$ & 121.9 & 340 & [312] & 1.460 & [1.46] & 0.01 & 0.78(0.20) & 1.8(0.5) & 1.26(0.32) & 0.86 \\
38 & Sr & 125 & 4s$^2$4p$^6$5s$^2$ & 229.4 & 1100 & [1050] & 2.375 & [2.375] & -0.24 & 0.88(0.14) & 1.9(0.7) & 1.19(0.22) & 0.89 \\
39 & Y & 125 & 4s$^2$4p$^6$5s$^2$4d$^1$ & 202.6 & 1850 & [1799] & 4.359 & [4.24] & -0.47 & 0.60(0.09) & 1.3(0.2) & 1.47(0.25) & 0.82 \\
40 & Zr & 125 & 4s$^2$4p$^6$5s$^2$4d$^2$ & 229.9 & 2500 & [2128] & 6.305 & [5.8] & 0.01 & 0.82(0.07) & 2.2(0.5) & 1.13(0.11) & 0.93 \\
41 & Nb & 125 & 4p$^6$5s$^1$4d$^4$ & 208.6 & 2900 & [2750] & 7.669 &  & 0.96 & 0.89(0.04) & 3.2(0.2) & 1.01(0.02) & 0.91 \\
42 & Mo & 125 & 5s$^1$4d$^5$ & 224.6 & 3000 & [2896] & 9.330 & [9.33] & 0.42 & 0.89(0.03) & 3.2(0.4) & 0.98(0.04) & 0.77 \\
43 & Tc & 125 & 5s$^2$4d$^5$ & 228.7 & 2500 & [2430] & 10.606 &  & -0.43 & 0.86(0.03) & 4.3(0.3) & 0.77(0.12) & 0.08 \\
44 & Ru & 125 & 5s$^1$4d$^7$ & 213.3 & 2800 & [2607] & 11.200 & [10.65] & 1.18 & 0.94(0.05) & 4.6(0.5) & 0.97(0.02) & 0.85 \\
45 & Rh & 125 & 5s$^1$4d$^8$ & 229.0 & 2400 & [2237] & 10.807 & [10.7] & -1.86 & 0.88(0.07) & 5.2(0.4) & 0.96(0.01) & 0.89 \\
46 & Pd & 125 & 5s$^1$4d$^9$ & 250.9 & 1900 & [1828] & 10.380 & [10.38] & -1.32 & 0.92(0.04) & 4.9(0.5) & 0.98(0.01) & 0.94 \\
47 & Ag & 125 & 5s$^1$4d$^{10}$ & 249.8 & 1350 & [1235] & 9.320 & [9.320] & 0.98 & 0.89(0.04) & 4.7(0.4) & 1.00(0.01) & 0.96 \\
48 & Cd & 125 & 5s$^2$4d$^{10}$ & 274.3 & 650 & [594] & 7.996 & [7.996] & 0.61 & 0.73(0.08) & 5.1(0.7) & 1.00(0.02) & 0.80 \\
49 & In & 125 & 5s$^2$5p$^1$ & 95.9 & 600 & [430] & 6.859 & [7.02] & 0.28 & 0.90(0.05) & 4.2(0.5) & 1.02(0.04) & 0.88 \\
50 & Sn & 125 & 5s$^2$5p$^2$ & 103.2 & 900 & [505] & 6.685 & [6.99] & -0.89 & 0.88(0.06) & 4.7(0.9) & 0.99(0.03) & 0.83 \\
51 & Sb & 125 & 5s$^2$5p$^3$ & 172.1 & 950 & [904] & 6.530 & [6.53] & 0.68 & 0.40(0.15) & 3.0(1.7) & 1.03(0.33) & 0.41 \\
52 & Te & 125 & 5s$^2$5p$^4$ & 175.0 & 750 & [723] & 5.700 & [5.70] & -0.64 & 0.14(0.30) & \multicolumn{1}{c|}{---} & \multicolumn{1}{c|}{---} & \multicolumn{1}{c|}{---} \\
\hline
55 & Cs & 125 & 5s$^2$5p$^6$6s$^1$ & 220.3 & 330 & [302] & 1.826 & [1.843] & 0.04 & 0.90(0.22) & 1.7(0.6) & 1.33(0.77) & 0.96 \\
56 & Ba & 125 & 5s$^2$5p$^6$6s$^2$ & 187.2 & 1050 & [1000] & 3.338 & [3.338] & -0.41 & 0.55(0.18) & 0.8(0.2) & 2.27(0.85) & 0.59 \\
57 & La & 125 & 5s$^2$5p$^6$6s$^2$5d$^1$ & 219.3 & 1280 & [1193] & 5.940 & [5.94] & -1.86 & 0.72(0.23) & 1.7(0.6) & 1.25(0.30) & 0.86 \\
72 & Hf & 125 & 5p$^6$6s$^2$5d$^2$ & 220.3 & 2600 & [2506] & 12.349 & [12] & -1.92 & 0.83(0.03) & 2.2(0.3) & 1.16(0.07) & 0.91 \\
73 & Ta & 125 & 5p$^6$6s$^2$5d$^3$ & 223.7 & 3450 & [3290] & 15.000 & [15] & 0.45 & 0.90(0.03) & 3.3(0.3) & 1.02(0.02) & 0.91 \\
74 & W & 125 & 5p$^6$6s$^2$5d$^4$ & 223.1 & 3900 & [3695] & 16.803 & [17.6] & -0.78 & 0.86(0.08) & 3.7(0.6) & 0.99(0.03) & 0.88 \\
75 & Re & 125 & 6s$^2$5d$^5$ & 226.2 & 3650 & [3459] & 18.787 & [18.9] & 0.98 & 0.82(0.09) & 4.5(0.7) & 0.97(0.02) & 0.81 \\
76 & Os & 125 & 6s$^2$5d$^6$ & 228.0 & 3450 & [3306] & 19.741 & [20] & 0.46 & 0.86(0.06) & 5.1(0.4) & 0.97(0.01) & 0.85 \\
77 & Ir & 125 & 6s$^1$5d$^8$ & 210.9 & 2900 & [2719] & 19.275 & [19] & -0.91 & 0.71(0.07) & 5.1(0.4) & 0.96(0.01) & 0.83 \\
78 & Pt & 125 & 6s$^1$5d$^9$ & 230.3 & 2200 & [2041] & 18.532 & [19.77] & -1.00 & 0.87(0.06) & 6.0(1.4) & 0.97(0.02) & 0.94 \\
79 & Au & 125 & 6s$^1$5d$^{10}$ & 229.9 & 1470 & [1337] & 16.690 & [17.31] & 1.06 & 0.86(0.14) & 7.9(1.6) & 0.95(0.02) & 0.92 \\
80 & Hg & 125 & 6s$^2$5d$^{10}$ & 233.2 & 470 & [234] & 11.917 &  & 0.48 & 0.84(0.13) & 4.4(1.9) & 1.00(0.07) & 0.67 \\
81 & Tl & 125 & 6s$^2$6p$^1$ & 90.1 & 600 & [577] & 11.220 & [11.22] & 0.53 & 0.90(0.09) & 3.7(0.8) & 1.06(0.04) & 0.86 \\
82 & Pb & 125 & 6s$^2$6p$^2$ & 98.0 & 900 & [601] & 10.671 & [10.66] & 0.48 & 0.90(0.10) & 4.1(0.7) & 1.03(0.03) & 0.94 \\
83 & Bi & 125 & 6s$^2$6p$^3$ & 105.0 & 580 & [545] & 10.050 & [10.05] & -0.27 & 0.17(0.02) & 1.6(1.5) & 1.40(0.13) & 0.14 \\
84 & Po & 125 & 6s$^2$6p$^4$ & 159.7 & 850 & [527] & 9.039 &  & -0.01 & 0.36(0.21) & 2.5(1.7) & 1.13(0.40) & 0.35 \\
\hline
\end{tabular}
  \caption{
    Correlation coefficients $R$ (Eq.\ (\ref{person})) and density scaling exponents $\gamma$  (Eq.\ (\ref{gammaWU})) 
    of the elements at the estimated DFT triple points $T,\rho$. 
    Values in round parenthesis are the statistical uncertainties on $R$ and $\gamma$.
    $N$ is the number of atoms in the periodic simulation cell.
    The electronic configuration of the calculated electrons as well as the
    plane wave cutoff energies $\eps_{\rm max}$ are listed. The last two columns gives the $\sigma$ parameter of the IPL approximation (Eq. \ref{ipl}) and the correlation between IPL and DFT energy fluctuations. Experimental values of $T$ and $\rho$ at the triple points are given in square brackets
    \cite{crc12,ass12,sav05,haa76}.
  }
  \label{tab:details}
\end{table*}
\endgroup

\section{Results}\label{results}

\subsection{Correlated virial and potential energy fluctuations}\label{ssec:p3}

\begin{figure} 
\begin{center} 
  \includegraphics[width=1.0\columnwidth]{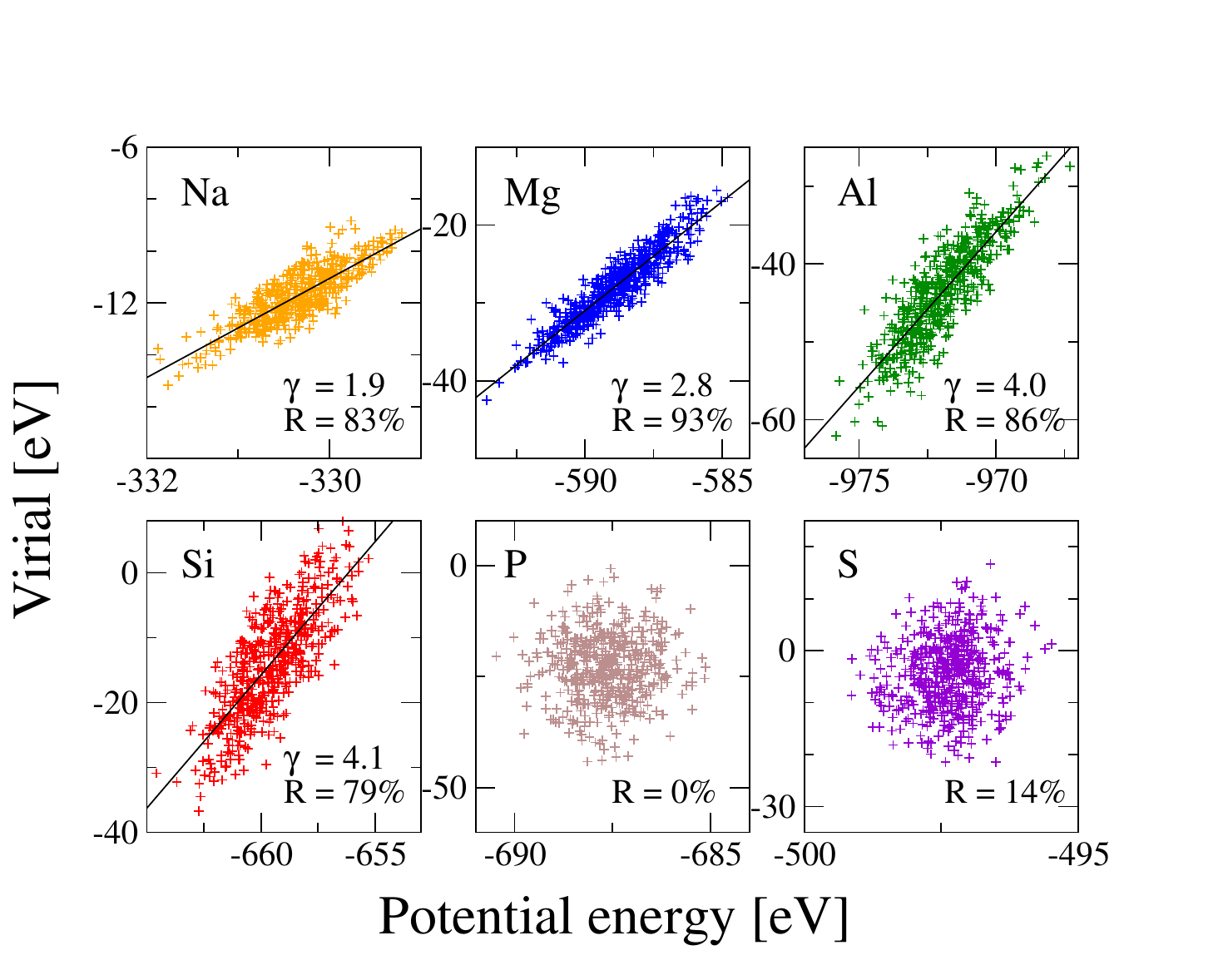}
\caption{\label{six_pack} Results from {\it ab initio} quantum-mechanical calculations of the first six period-three elements in the liquid phase at the triple point. Each subfigure shows a scatter plot of the potential energy $U(\bR,V)$ and the corresponding virial $W(\bR,V)$ for configurations of the $NVT$ ensemble. Fluctuations of virial and potential energy are strongly correlated for the metals Na, Mg and Al and the metalloid silicon Si, but not for the covalently bonded non-metals P and S. The strong correlation validates hidden scale invariance for metals. The subfigures report the Pearson correlation coefficient $R$ (Eq.\ (\ref{person})) and the density-scaling exponent $\gamma$ of Eq.\ (\ref{WUFluc}) determined as the linear-regression slope (Eq.\ (\ref{gammaWU})).}
\end{center} 
\end{figure}

Fig. \ref{six_pack} shows the results of DFT computations on the first six period-three elements. Each subfigure gives a scatter plot of virial versus potential energy of configurations taken from $NVT$ equilibrium simulations of the liquid phase at the triple points. For the metals (Na, Mg and Al) and the metalloid (Si) the scatter plots show strong correlations, implying that Eq.\ (\ref{WUFluc}) is obeyed to a good approximation. The value of the Pearson correlation coefficient $R$ (Eq.\ (\ref{person})) quantifies how well the approximation is obeyed \cite{ped08,I,II,III,IV}. The non-metals P and S do not exhibit strong $WU$ correlations.

\begin{figure} 
\begin{center} 
  \includegraphics[width=0.45\columnwidth]{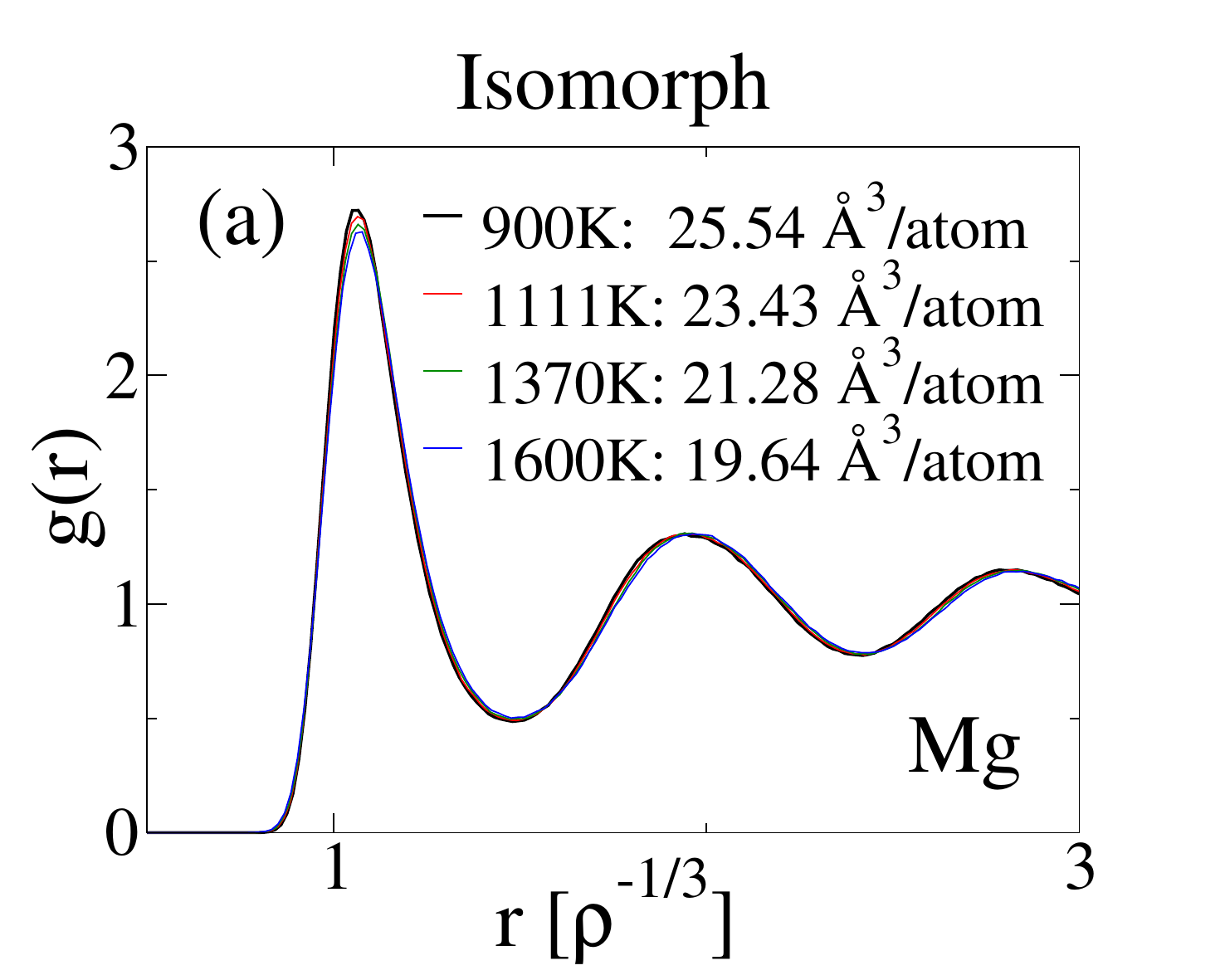}
  \includegraphics[width=0.45\columnwidth]{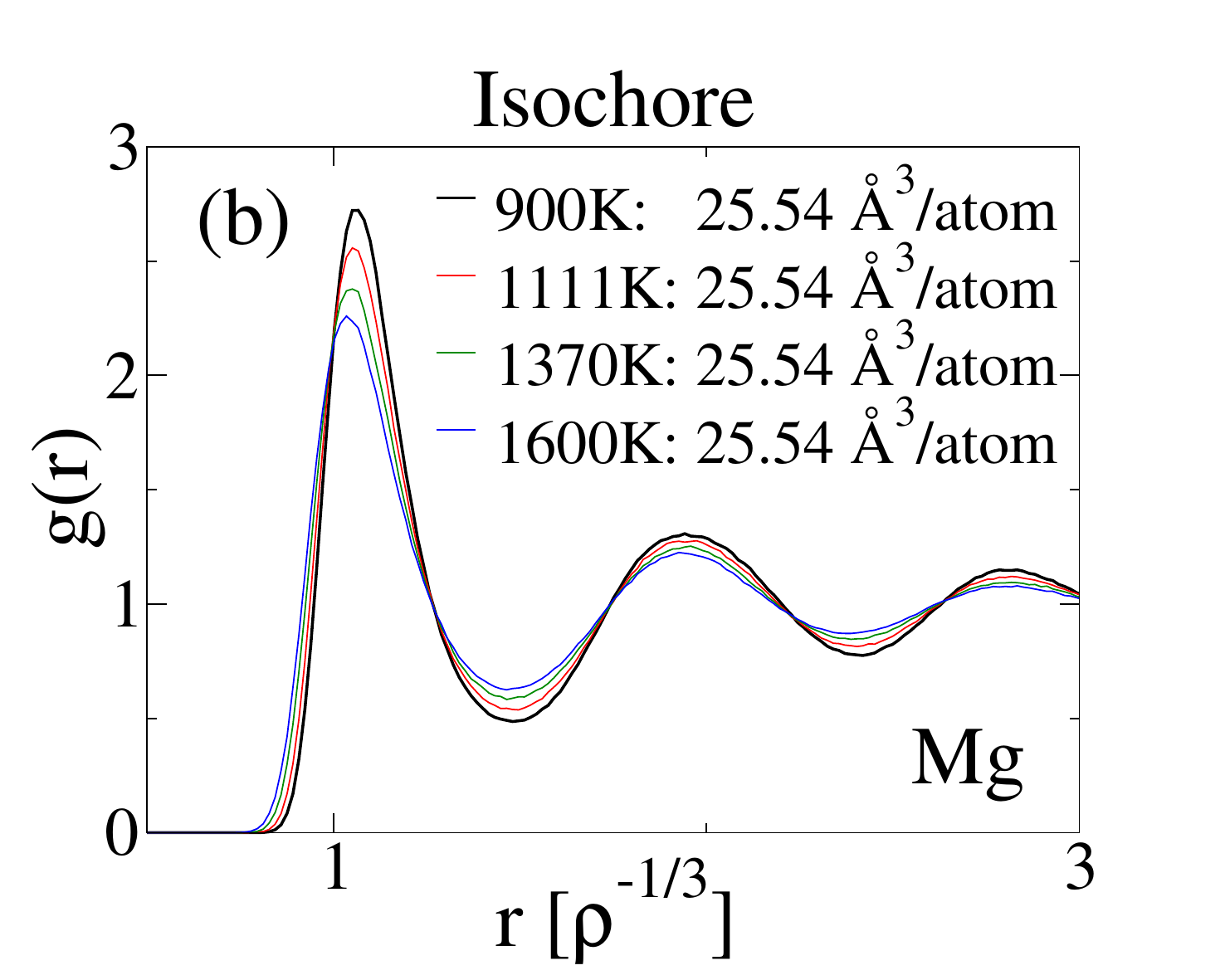} \\
  \includegraphics[width=0.45\columnwidth]{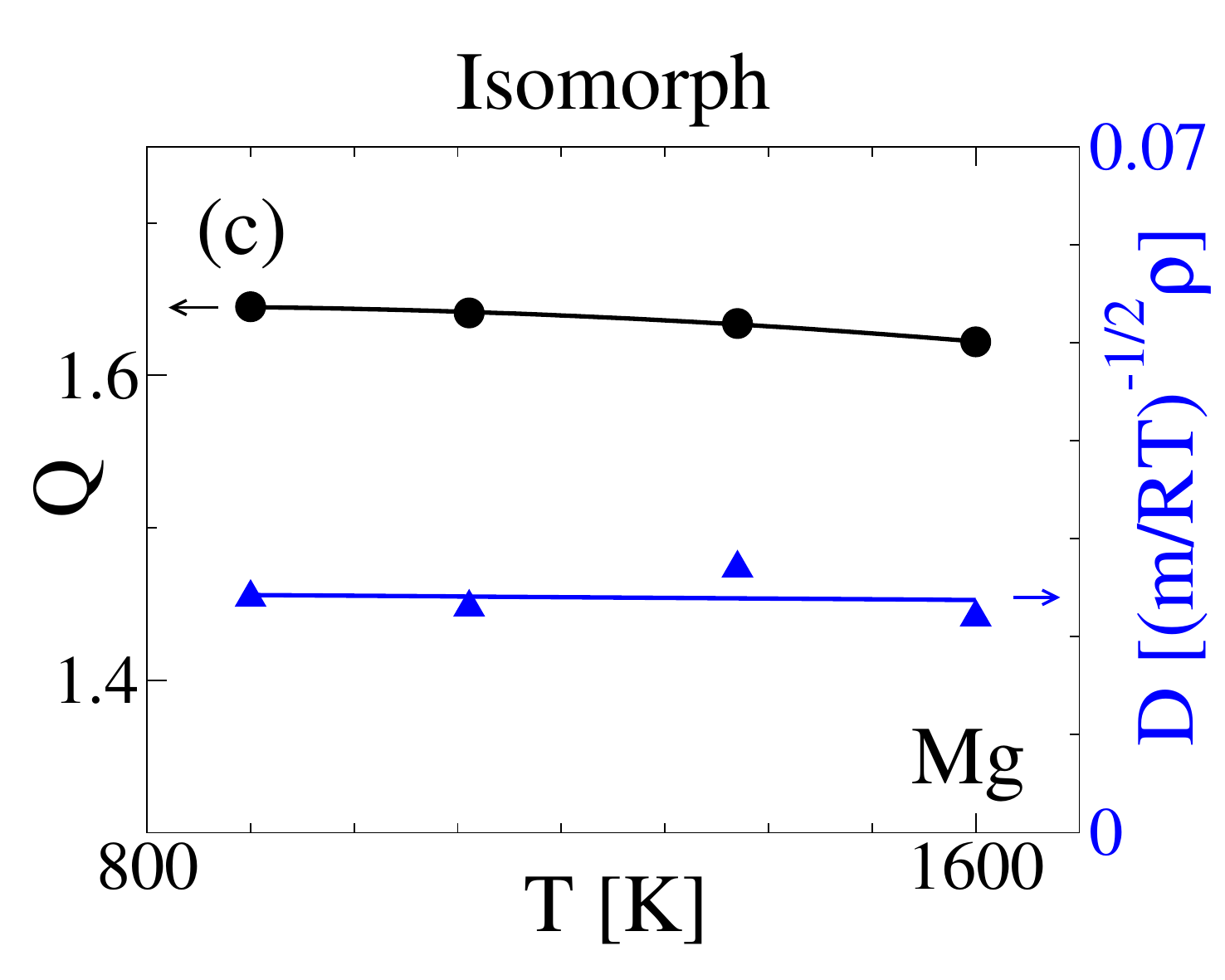} 
  \includegraphics[width=0.45\columnwidth]{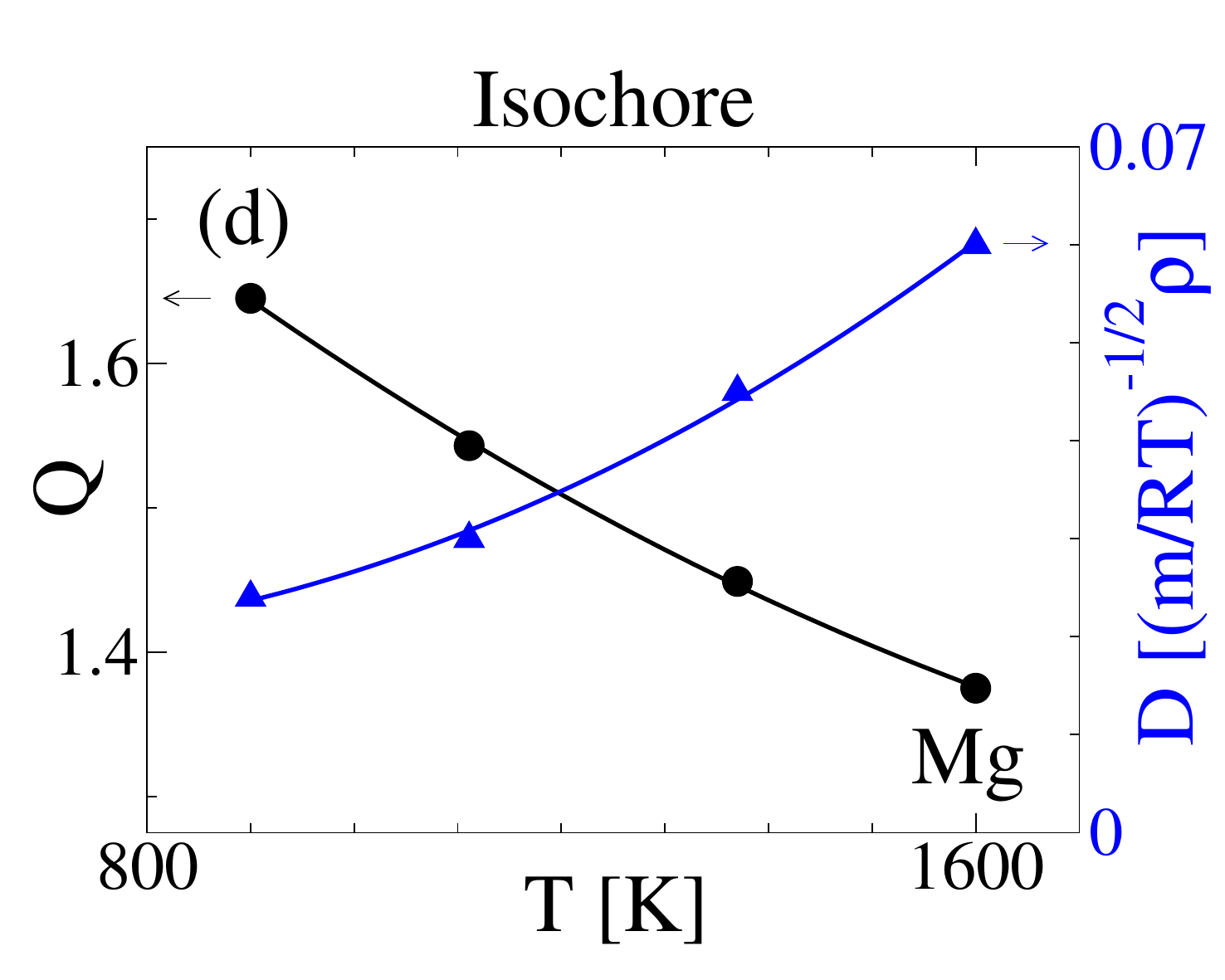}
\caption{\label{isom} Predicted isomorph invariants of magnesium along an isomorph and an isochore for the same temperature variation. Panels (a) and (b) show results for the radial distribution function $g(r)$ in reduced units, panels (c) and (d) show the translational order parameter $Q$ of Debenedetti {\it et al.} \cite{tru00} and the reduced diffusion constant $D$. The reduced unit structural and dynamical quantities $g(r)$, $Q$, and $D$ all vary much less along the isomorph ((a) and (c)) than along the isochore ((b) and (d)). }
\end{center}
\end{figure}

\subsection{Invariant structure and dynamics}
To demonstrate hidden scale invariance directly, we investigated liquid magnesium in more detail. The results are summarized in Fig. \ref{isom}, which studies isomorphic as well as isochoric state points for temperatures between 900 K and 1600 K. Panels (a) and (b) give the radial distribution functions for the isomorphic and isochoric state points, respectively, while (c) and (d) give the translational order parameter $Q$ suggested by Truskett, Torquato and Debenedetti \cite{tru00} and the reduced diffusion constant. We see that structure and dynamics are almost invariant along the isomorph, which confirms magnesium's hidden scale invariance.

\begin{figure}
  \includegraphics[width=0.55\columnwidth]{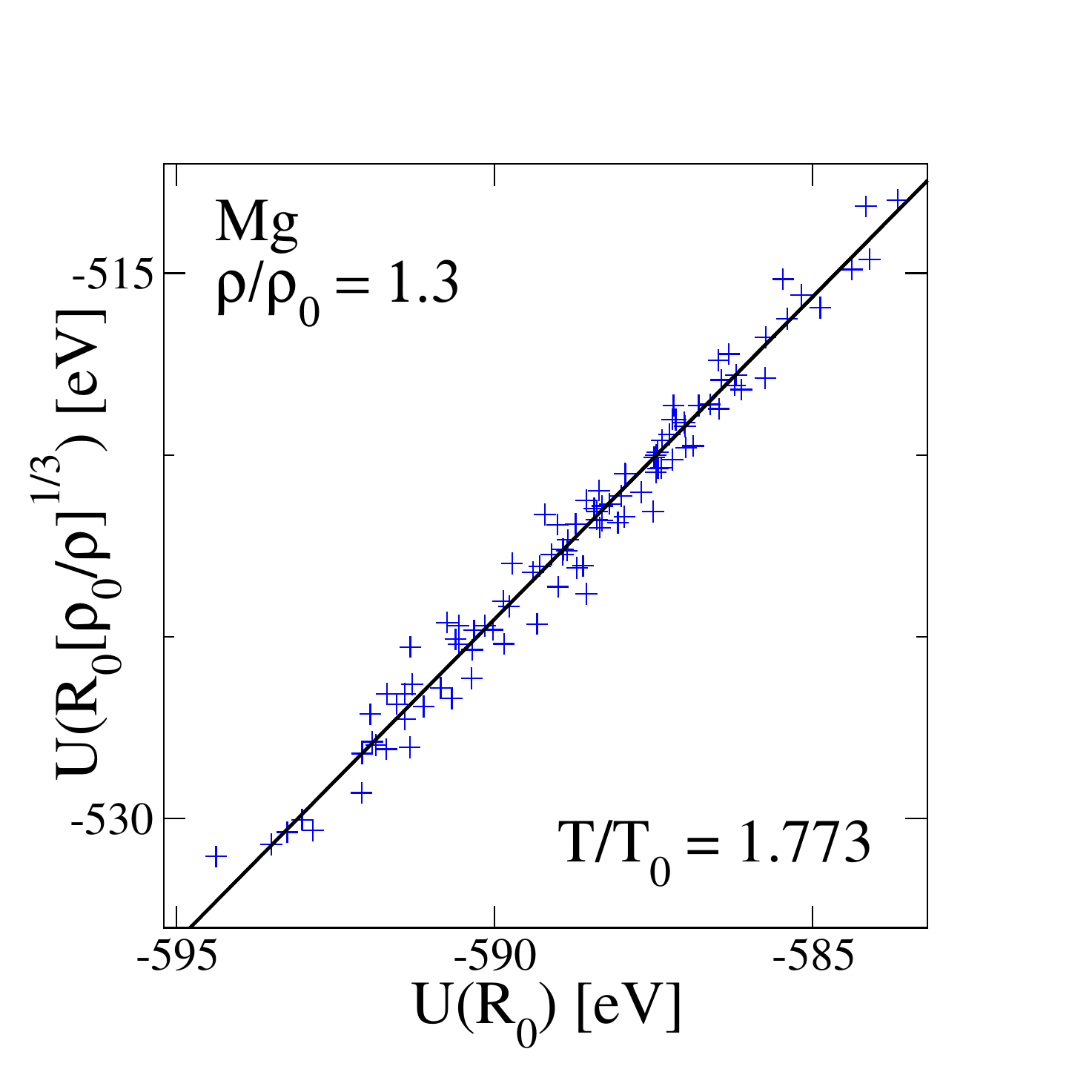}
  \caption[Calculating isomorphic temperatures]{
    Determination of the isomorphic temperatures at $\rho/\rho_0=1.3$ for Mg using Eq.\ (\ref{boltzmann}) (with $\rho_0=1.581$ g/cm$^3$ being the triple point density). The ratio $T/T_0$
    is given as the ratio of the fluctuations of the inner energy of the scaled and unscaled configurations.
  }
  \label{fig:isomorphTemp}
\end{figure}

The isomorph in Fig \ref{isom} is determined as follows: First we choose to identify state points with density increases
of $9\%$, $20\%$ and $30\%$ relative to that of the triple point. Then temperatures along the isomorph are determined by relating the Boltzmann factors of scaled configurations \cite{III}:
\begin{equation}\label{boltzmann}
  \exp({-U(\vec R_0)/k_BT_0}) \propto \exp({-U(\vec R)/k_BT}).  
\end{equation}
For 100 representative configurations $\vec R_0$ at the reference point (i.e. the triple point) $(T_0,\rho_0)$, configurations are rescaled to the new density $\rho$ and the energy of the scale configuration $U(\vec R) = U(\vec R_0 [\rho_0/\rho]^{1/3})$ is evaluated with DFT. By taking the logarithm of Eq.\ (\ref{boltzmann}), we see that $T/T_0$ is the ratio of the fluctuations of the energy at the reference point $U(\vec R)$ and the energy of the scaled configurations $U(\vec R_0)$. Fig.\ \ref{fig:isomorphTemp} show an example of a scatter plot of these. The computed temperatures of the isomorphic states are 1110\,K, 1370\,K and 1600\,K, respectively.

\subsection{The periodic table of hidden scale invariance}
\begin{sidewaysfigure*}
\vspace{10cm}
\begin{center} 
\includegraphics[width=1.0\columnwidth]{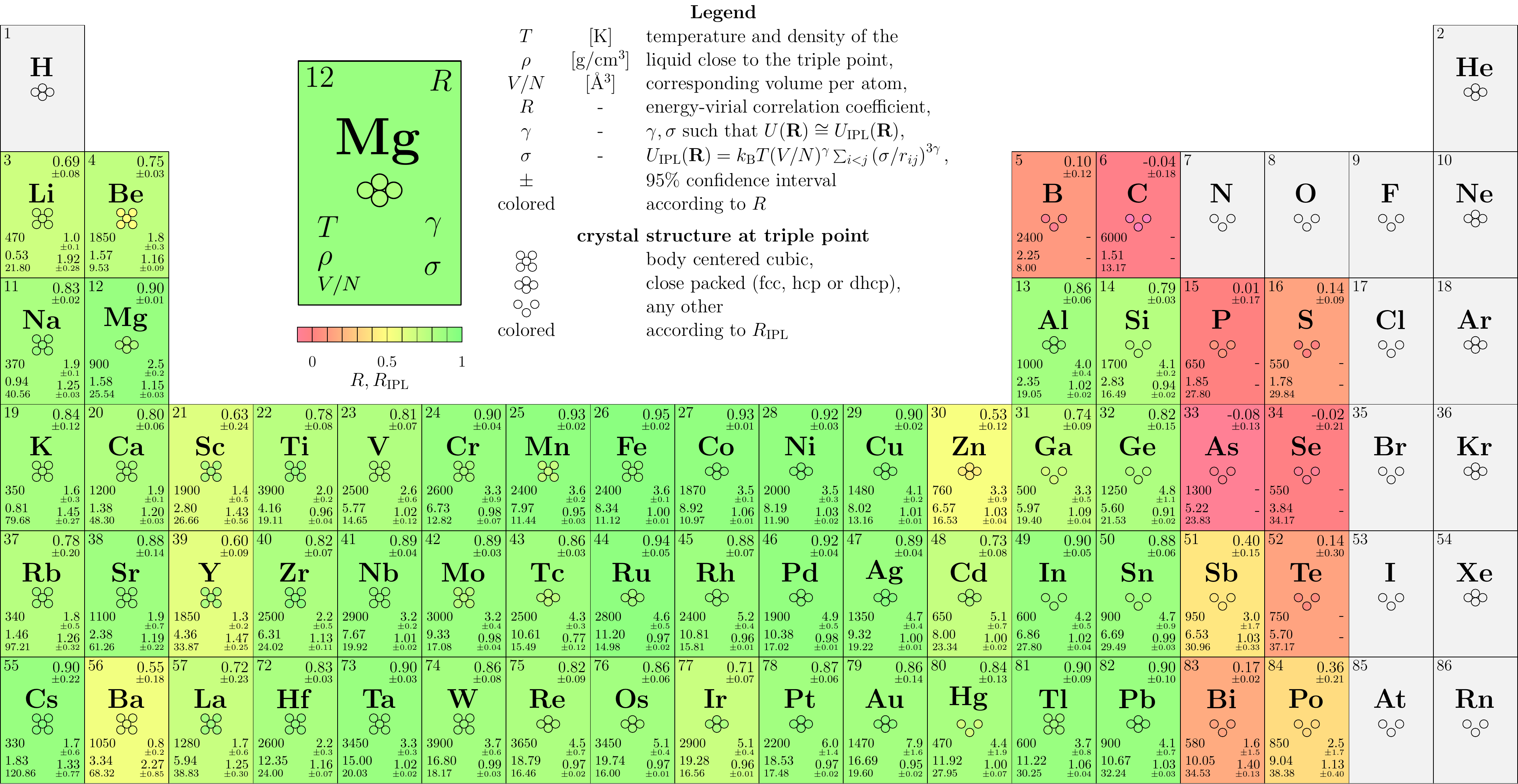}
\caption{\label{ptable} Periodic table of elements with color coded bacgrounds indicating the virial potential-energy correlation coefficient $R$. Greenish colors correspond to elements with strong correlations between fluctuations of virial and potential energy in the $NVT$ ensemble of the liquid state at the triple point. These elements exhibit hidden scale invariance. The thick line separates metals and non-metals. Most metalloids and all non-metals investigated have weak to vanishing correlations. Gray elements were not considered since standard semi-local density functionals are inaccurate for these. See legend and Table \ref{tab:details} for more details. 
}
\end{center} 
\end{sidewaysfigure*}

The results of Fig.\ \ref{six_pack} inspired us to study the elements in general to investigate whether all liquid metals exhibit hidden scale invariance. The results for 58 elements are summarized in Fig.\ \ref{ptable} and Table \ref{tab:details}. We excluded some non-metallic elements (gray on Fig.\ \ref{ptable}) for which standard semi-local density functionals are inaccurate. 
Metallic liquid elements all have strong or fairly strong virial potential-energy correlations at the triple point as quantified in the virial potential-energy correlation coefficient $R$. 
Most of the metals have a $R$ larger than 80\%, however, a few metallic elements show correlation in the range of 50\%-60\% (Li, Sc, Zn, Y, Ba). Scale invariance is expected to be worse for these elements.

As mentioned, all systems have the hidden-scale-invariance property at high pressure where repulsive pair interactions dominate (see Sec. \ref{ssc:HighPressure}). Moreover, crystals generally have stronger virial potential-energy correlations than liquids \cite{alb14}. We therefore conclude that metallic elements are (R) simple in the entire condensed-phase part of the phase diagram, i.e., exhibit hidden scale invariance. This excludes state points close to the critical point, as well as those of the gas phase far from the melting line.

Table \ref{tab:details} also reports the computed DFT scaling exponents $\gamma$, which we discuss in details in the following Sec. \ref{subsection:gamma}.

\subsection{Density scaling exponents}\label{subsection:gamma}
\begin{figure} 
\begin{center} 
  \includegraphics[width=0.80\columnwidth]{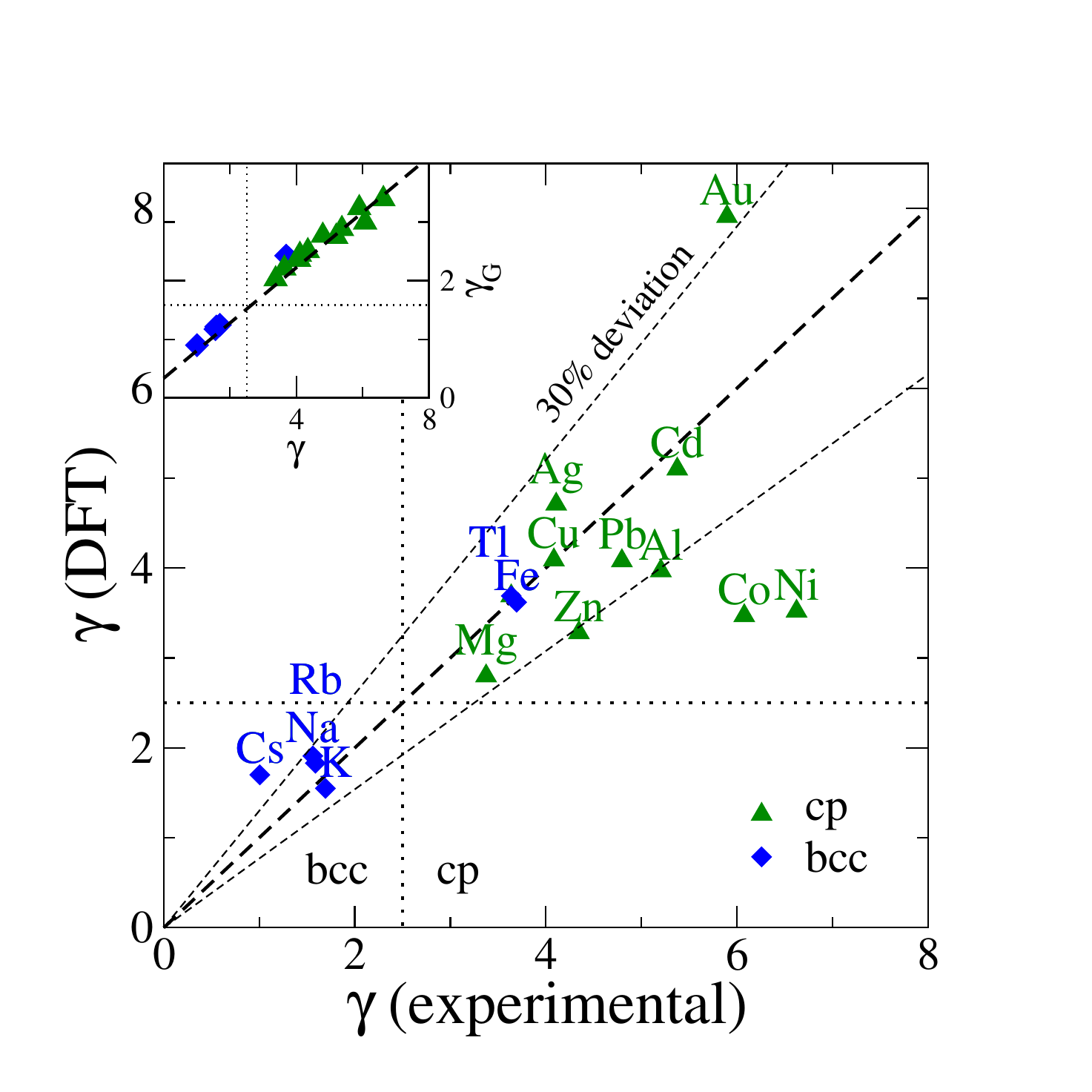} 
\end{center}
\caption{\label{DFTCompare} Density scaling exponents from experimental thermodynamic data collected in Ref. \cite{sin07} compared to {\it ab initio} DFT computations for liquids. Elements forming bcc structures are marked with blue diamonds, while elements forming one of the close-packed structures (fcc or hcp) are marked with green triangles. 
The inset compares the experimental $\gamma$'s to the Gr{\"u}neisen parameter $\gamma_G$. The dashed line indicates the Dulong-Petit approximation ($c_v\simeq3k_B$).}
\end{figure}

For a monatomic liquid of $N$ classical particles (above the Debye temperature) the density scaling exponent $\gamma$ can be determined purely from thermodynamic responses \cite{III}: 
\begin{equation}\label{exponent} 
\gamma \,=[\gamma_G-k_B/c_v]/[1-3k_B/2c_v]\,. 
\end{equation}
where
 $\gamma_G=\alpha_p K_T/\rho c_v$
is the thermodynamic Gr{\"u}neisen parameter. Here, $\alpha_p=[\partial V/\partial T]_p/V$ is the isobaric thermal expansion coefficient, $K_T=-V[\partial p/\partial V]_T$ is the isothermal bulk modulus, $c_v=C_V/N$ is the isochoric heat capacity per atom, and $k_B$ is the Boltzmann constant. The density scaling exponent is proportional to the thermodynamic Gr{\"u}neisen parameter up to the accuracy of the Dulong-Petit approximation $c_v\simeq3k_B$: $\gamma\simeq2\gamma_G-2/3$ (see inset on Fig.\ \ref{DFTCompare}). In Fig.\ \ref{DFTCompare} we compare 16 experimentally determined density scaling exponents \cite{sin07} to the values computed with DFT for the liquid state (see Tab. \ref{tab:details}).
Agreement is generally quite good.

Incidentally, our computations show that the acclaimed Lennard-Jones model does not properly reflect the physics of most metals because the DFT values of the density-scaling exponents are generally significantly smaller than $\gamma\simeq6$ of the Lennard-Jones model \cite{ped08}.

Interestingly, with the exception of Fe and Tl, the elements Na, K, Rb and Cs forming the more open body centered cubic (bcc) crystal structure have lower scaling exponents than the elements Mg, Al, Co, Ni, Cu, Zn, Ag, Cd, Au and Pb that form close packed (cp) structures -- face centered cubic (fcc) or hexagonal close packed (hcp). We will explore this further in the following Sec. \ref{subsection:ipl}.

\subsection{Revisiting the inverse power-law model}\label{subsection:ipl}

\begin{figure} 
\begin{center} 
  \includegraphics[width=0.80\columnwidth]{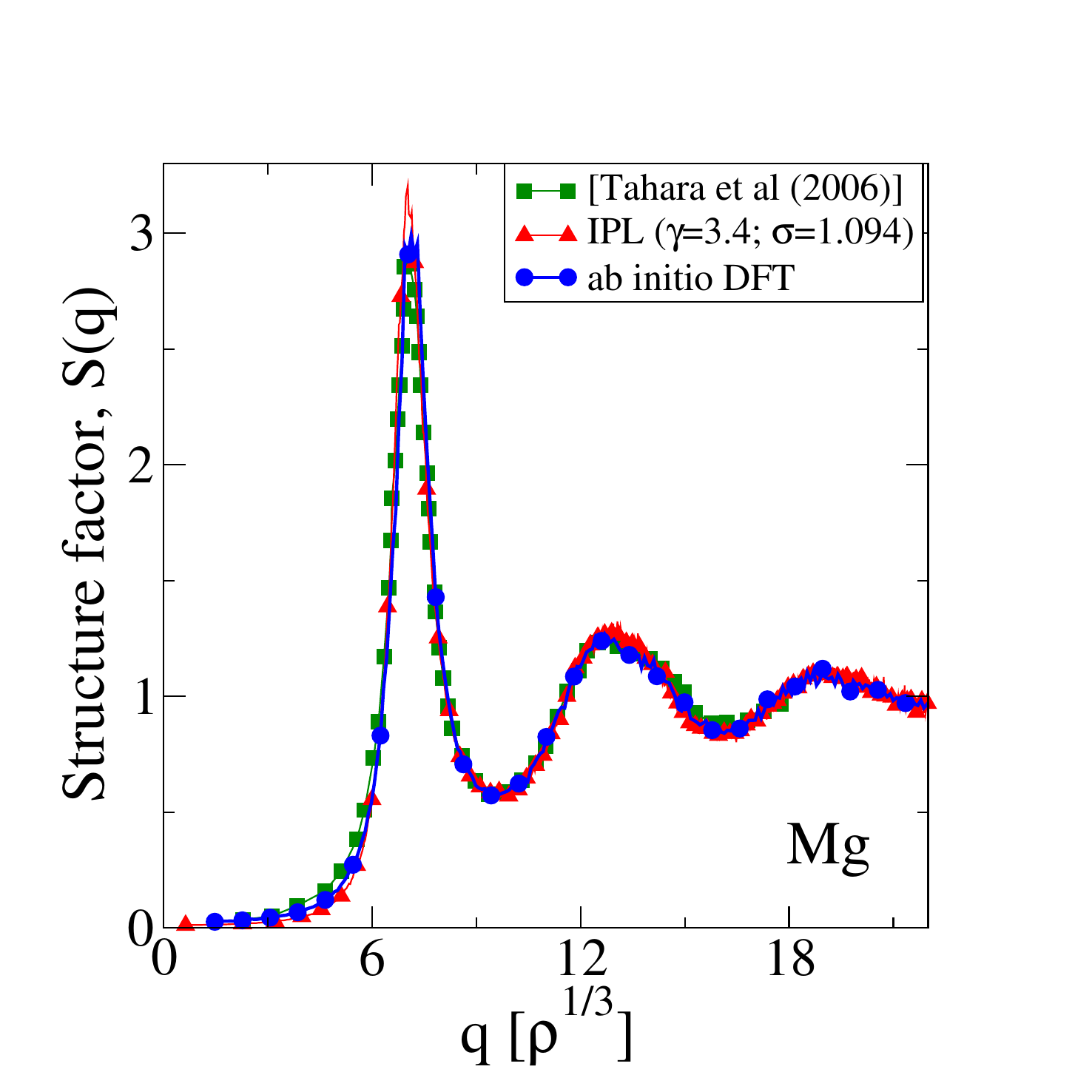}
\end{center}
\caption{\label{FigSq} Experimental structure factor of Mg at the triple point \cite{tah06} (green squares), compared to the IPL prediction (red triangles). This prediction has no free parameters: The exponent $\gamma=3.4$ of the IPL potential (Eq.\ \ref{ipl}) is given by experimental values \cite{sin07} of $c_v$, $\alpha_p$ and $K_T$ (using Eq.\ (\ref{exponent})), and $\sigma=1.094$ was chosen to match that of the solid-liquid coexistence shown in Fig.\ \ref{figIPL}. The blue dots are the results of an {\it ab initio} DFT calculation. The agreement is excellent.}
\end{figure}

As mentioned in the introduction, hidden scale invariance may be explained if the interactions in a metal can be approximated with the IPL model \cite{hoo71,hoo72,hiw74,sti75,ros76,you77,ros83,you91,pre05,bra06,hey07,hey08,ped10,bra11,khr11a,tra14}: $U(\bR)\simeq U_\textrm{IPL}(\bR)$ where
\begin{equation}\label{ipl} 
U_{\rm IPL}(\bR)= Ng(\rho)+A\sum_{i>j}^N |\br_{j}-\br_{i}|^{-3\gamma}
\end{equation}
where $A=k_BT\sigma^{3\gamma}\rho^{-\gamma}$ is a materials dependent constant. The mean field term $Ng(\rho)$ takes long-ranged attractive interactions into account \cite{waals1873,wid67}. Scale invariance is not ``hidden'' for the IPL model, but a trivial result: Temperature and density merge into the dimensionless parameter $\sigma=[A\rho^\gamma/k_BT]^{1/3\gamma}$ -- i.e. $\sigma$ is the single parameter of the phase diagram and isomorphs are given by $T\propto\rho^{-\gamma}$.
%
To quantify the accuracy this approximation we list the correlation between the IPL energy and the DFT energy,
$
  R_{\rm IPL} = \langle \Delta U_{\rm IPL}(\bR) \Delta U(\bR) \rangle/\sqrt{\langle [\Delta U_{\rm IPL}(\bR)]^2 \rangle\langle [\Delta U(\bR)]^2 \rangle},
$
in Table \ref{tab:details} and color code symbols indicating the crystal structures in Fig. \ref{ptable} accordingly. Elements with strong correlations between $W$ and $U$ fluctuations also have strong correlations between $U_{\rm IPL}$ and $U$. 
Fig.\ \ref{FigSq} compares the experimentally determined structure factor $S(q)$ of Mg at the triple point \cite{her99,tah06,sen09} with that of the IPL model (without using any free parameters). The agreement is excellent. The deviation at short $q$ vectors, related to the bulk modulus, is due to the mean-field $g(\rho)$ treatment of long-ranged attractive interactions \cite{mcl82}.

\begin{figure} 
\begin{center} 
  \includegraphics[width=0.80\columnwidth]{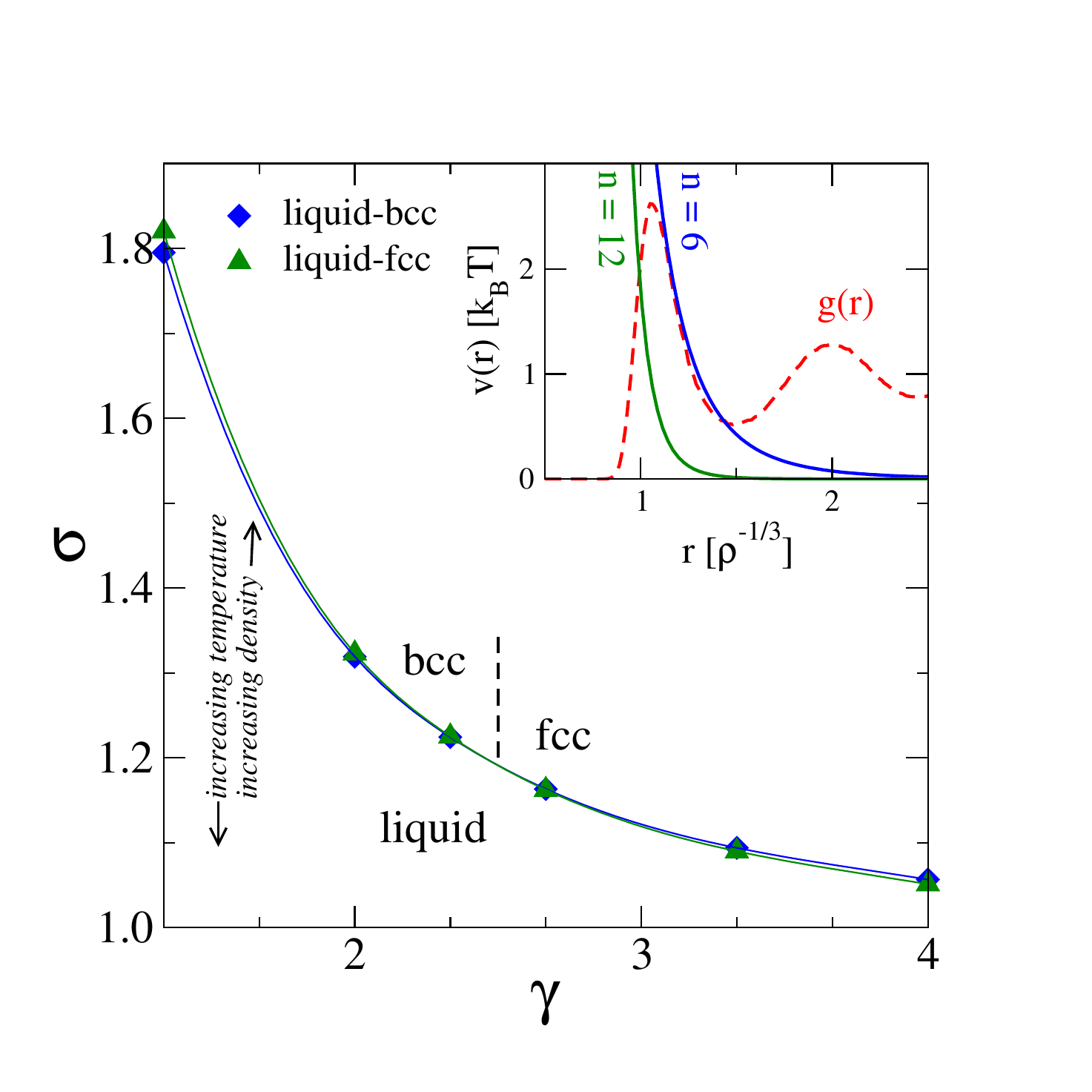}
\end{center}
\caption{\label{figIPL}Phase diagram of the IPL model (Eq.\ (\ref{ipl}) with $k_BT=\rho=1$ and $g(\rho)=0$) computed with the interface pinning method \cite{ped13}. Green triangles are the liquid sides of the liquid-fcc transitions, and blue diamonds are liquid-bcc transitions. Full lines are guides to the eye. The vertical dashed line indicate the bcc-fcc transition at $\gamma=2.5$ \cite{hoo72,agr95}. The inset show the pair energies and a representative liquid radial distribution at coexistence.
}
\end{figure}

\begin{figure} 
\begin{center} 
  \includegraphics[width=1.0\columnwidth]{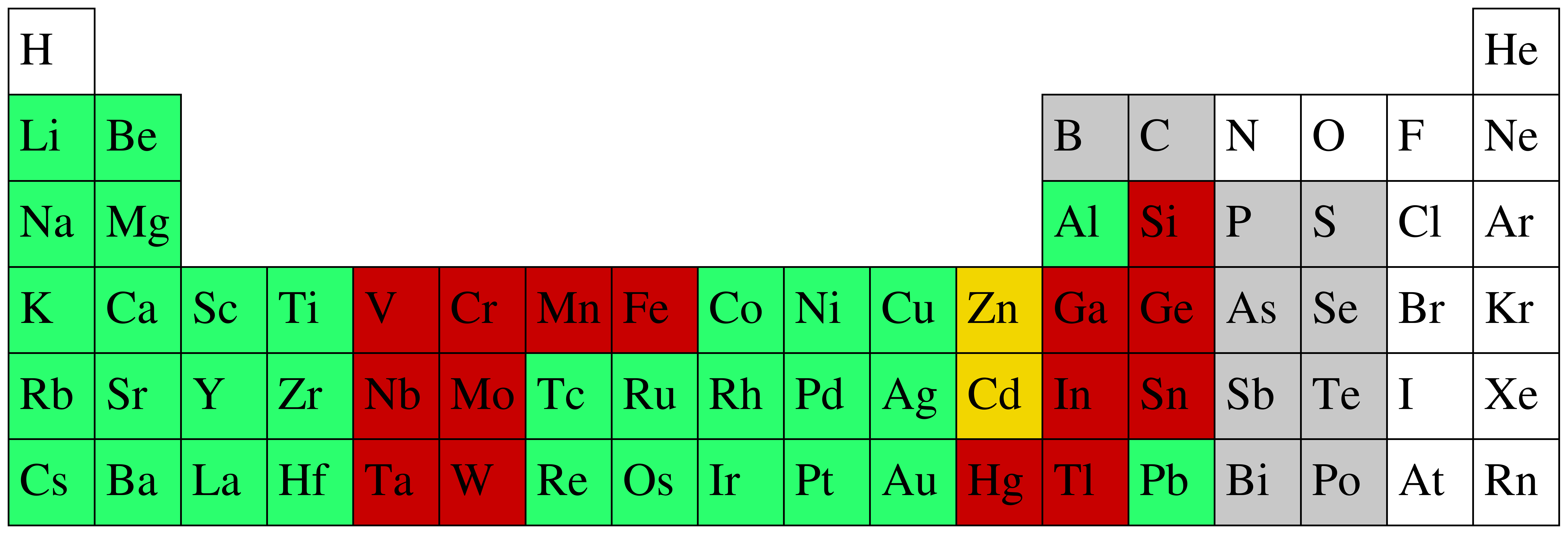}
\end{center}
\caption{\label{ptCrystal} Periodic table where green elements has a crystal structure at the triple point \cite{you91,ton04,gri12} predicted by assuming IPL interactions (see Sec. \ref{subsection:ipl}). The IPL prediction fails for red elements. Zn and Cd have been colored yellow since their distorted hcp crystals makes the IPL prediction ambiguous. All colored elements have hidden scale invariance seen in correlated $WU$ fluctuations (see Fig.\ \ref{ptable} and Table \ref{tab:details}).}
\end{figure}

Inspired by the seminal 1972 paper by Hoover, Young and Grover \cite{hoo72} we look for a connection between density scaling exponents and crystal packings: An IPL liquid is known \cite{hoo71,hoo72,you91,lai92,agr95,pre05,khr11a} to crystallize into a close packed (cp) fcc crystal structure when the IPL $\gamma$ is above 2.5, while the more open body centered cubic (bcc) crystal is stable at lower $\gamma$'s (Fig.\ \ref{figIPL}). 
The IPL model exhibit a polymorphic cp-bcc transition when $\gamma<2.5$ also seen for many metals \cite{you91,ton04,gri12}.
Thus, at the triple point elements with $\gamma<2.5$ are expected \cite{hoo72} to form bcc crystals while elements with $\gamma>2.5$ are expected to form cp crystals (fcc or hcp).
Fig.\ \ref{ptCrystal} shows the agreement of this prediction with the experimental crystal structure formed at the triple point (using the DFT $\gamma$'s given in Table \ref{tab:details}). 
Interestingly, the predictions fails for several transition metals (V, Cr, Mn, Fe, Nb, Mo, Ta, W and Hg), most post-transition metals (Ga, In, Sn and Tl) and the metalloids Si and Ge. Moreover, the IPL model predicts that metals that form bcc structures at the triple point also have a low temperature cp phase -- this is not always the case.
For all of these elements we find significant $WU$ correlations (see Fig.\ \ref{ptable}).
Thus, we conclude that scale invariance can be present even when the elements crystal structure cannot be accurately predicted by IPL-like interactions. This also illustrates the importance of choosing an {\it ab initio} method for investigating hidden scale invariance.

\subsection{The melting lines follow isomorphs}

The melting line follows an isomorph to a good approximation \cite{III,PhysRevB.88.094101,IV}. In Table \ref{tbl} this prediction is validated for the period-three metals by showing that the melting-line scaling exponent $\gamma_\textnormal{m}\equiv d \ln T_{\rm m}/d\ln\rho$ \cite{V,alb14,ped13,PhysRevB.88.094101} agrees with that of the isomorph, the density-scaling exponent $\gamma$.

\begin{table} 
  \caption{\label{tbl}Correlation coefficients and scaling exponents for the period-three metals in the liquid phase at the triple point. Comparing the last two columns show that the melting-line scaling exponent $\gamma_m$ agrees with the density-scaling exponent $\gamma$. }
\begin{ruledtabular} 
    \begin{tabular}{lcccc} 
      Element  &  $R$ & $\gamma$ & $\gamma_\textnormal{m}$ \\ \hline
      Na  & $0.83(0.02)^a$ & $1.9(0.1)$ & $1.7(0.5)$ \\
      Mg  & $0.93(0.04)$   & $2.8(0.3)$ & $2.6(0.5)$   \\
      Al  & $0.86(0.06)$   & $4.0(0.4)$ & $4.6(0.8)$   \\
    \end{tabular}
 \end{ruledtabular}
\begin{flushleft}
    $^a$Numbers in parenthesis indicate the statistical uncertainties. 
\end{flushleft}
\end{table}

\subsection{Elevated pressure behavior of Fe and P}
\label{ssc:HighPressure}

Fe shows a fairly good correlation between the virial and the potential energy.
Interestingly, 
the melting line $T_m(\rho)$ reported in Refs.  \cite{poirier,PhysRevB.87.094102} coincides to a good degree with an extrapolation along the isomorph starting from the triple point, using $T_m\propto\rho^\gamma$ with $\gamma=3.6$ (see Table \ref{tab:details}).

\begin{figure*} 
\begin{center} 
\includegraphics[width=0.30\textwidth]{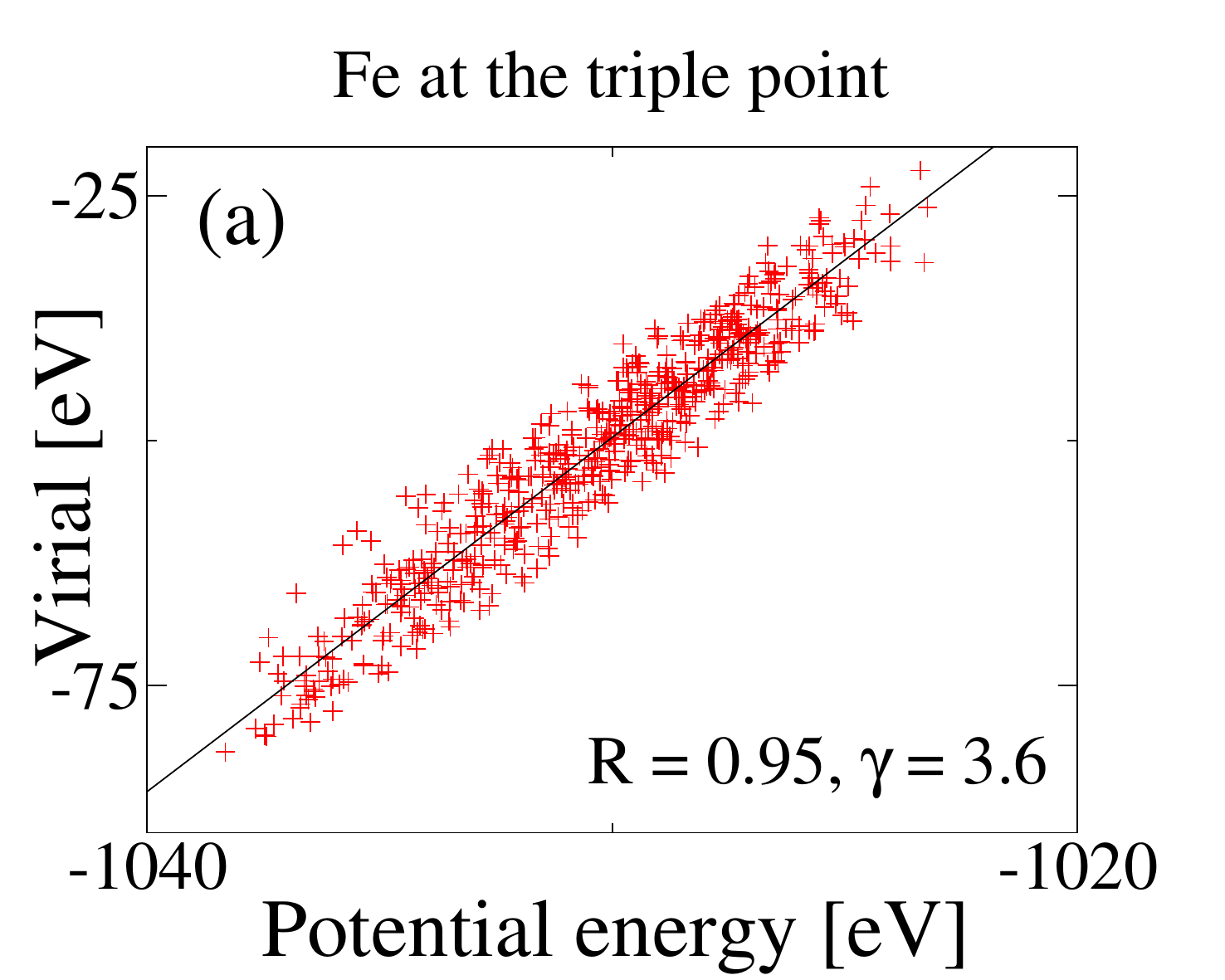} 
\includegraphics[width=0.30\textwidth]{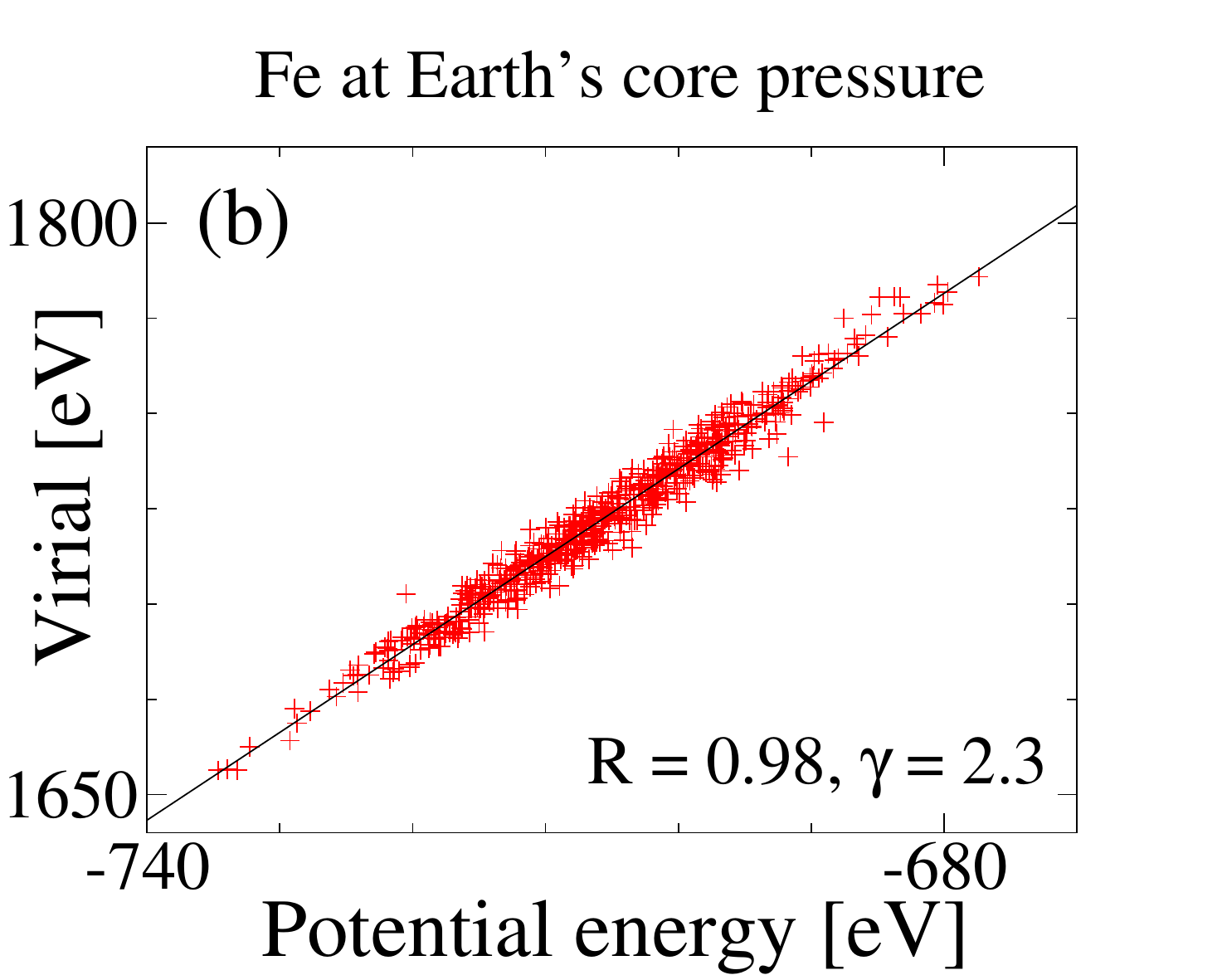} 
\includegraphics[width=0.30\textwidth]{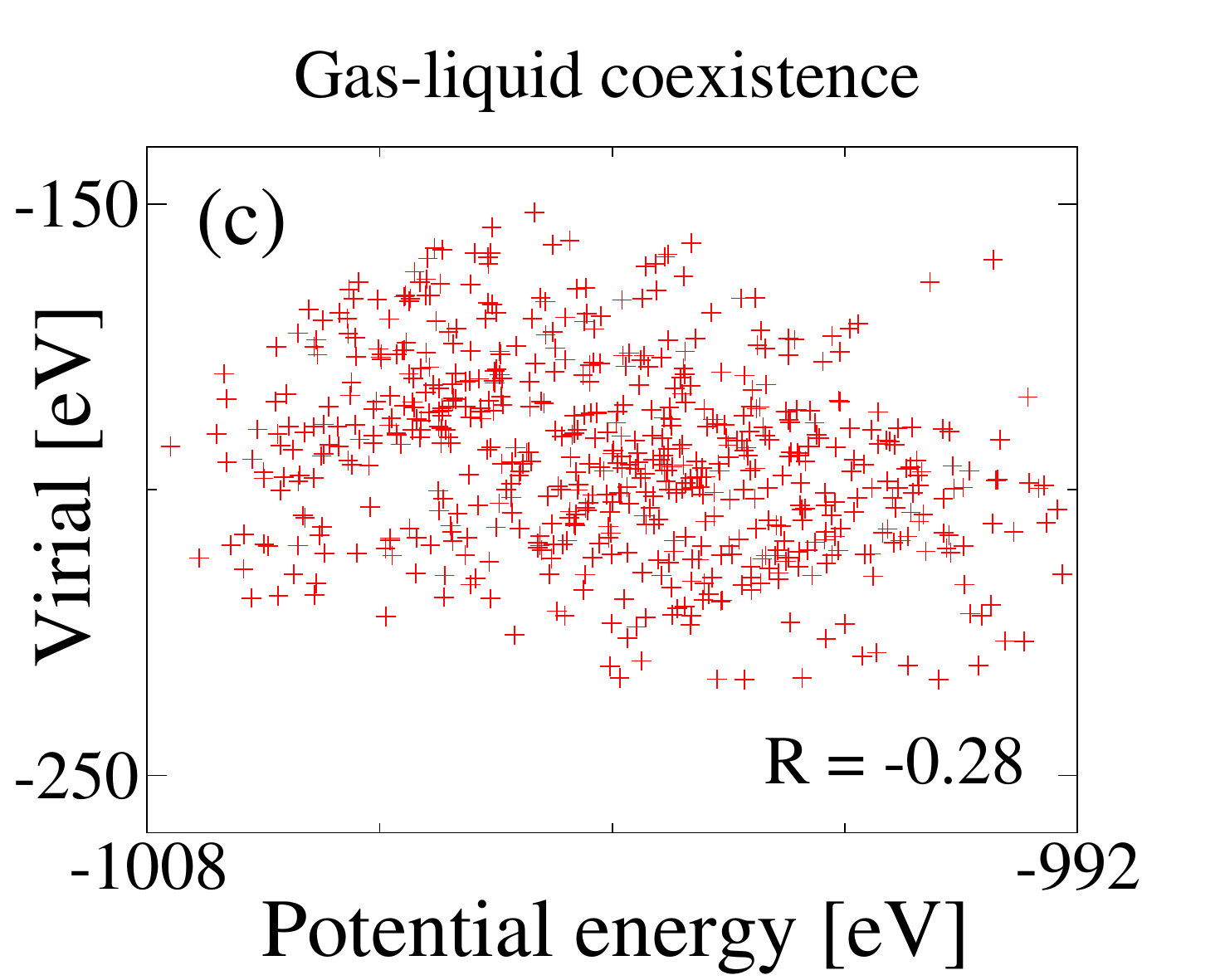}
\caption{\label{fe} Virial potential-energy scatter plots for iron.
Panel (a) shows a scatter plot of potential energy and virial of 125 iron atoms at the DFT triple point. Panel (b) shows that the correlation increases at elevated pressure, here chosen corresponding to a pressure at the Earth's core and near the melting line of pure iron (310 GPa, 9000 K) \cite{poirier,PhysRevB.87.094102}. Thus iron becomes simpler at higher pressures. Panel (c) shows that the correlation is low in the gas-liquid coexistence region.  
}
\end{center}
\end{figure*}

To exemplify that the behavior of elements becomes simpler at high pressure we compare in Fig.\ \ref{fe} the $WU$ scatter plot of Fe at the triple point (Fig.\ \ref{fe}(a)) with simulation results at the pressure 310 GPa (Fig.\ \ref{fe}(b)) corresponding to a pressure in Earth's liquid outer core, and near the melting line of pure iron \cite{poirier,PhysRevB.87.094102}.  The correlation coefficient increases from $R=0.95$ to $R=0.98$.  Fig.\ \ref{fe}(c) shows that $WU$ fluctuations are uncorrelated in the gas-liquid coexistence regime, which is presumably associated with the formation of a gas-liquid interface in the simulation cell and the onset of critical fluctuations. 
Likewise, the $WU$ correlation coefficient of the non-metal phosphor (P) increases at high pressures from nearly uncorrelated at the triple point to $R=0.30$ and $R=0.52$ at ($p=63$ GPa, $T=1880$ K) and ($p=210$ GPa, $T=2200$ K), respectively.

\begin{figure}
  \includegraphics[height=0.45\columnwidth]{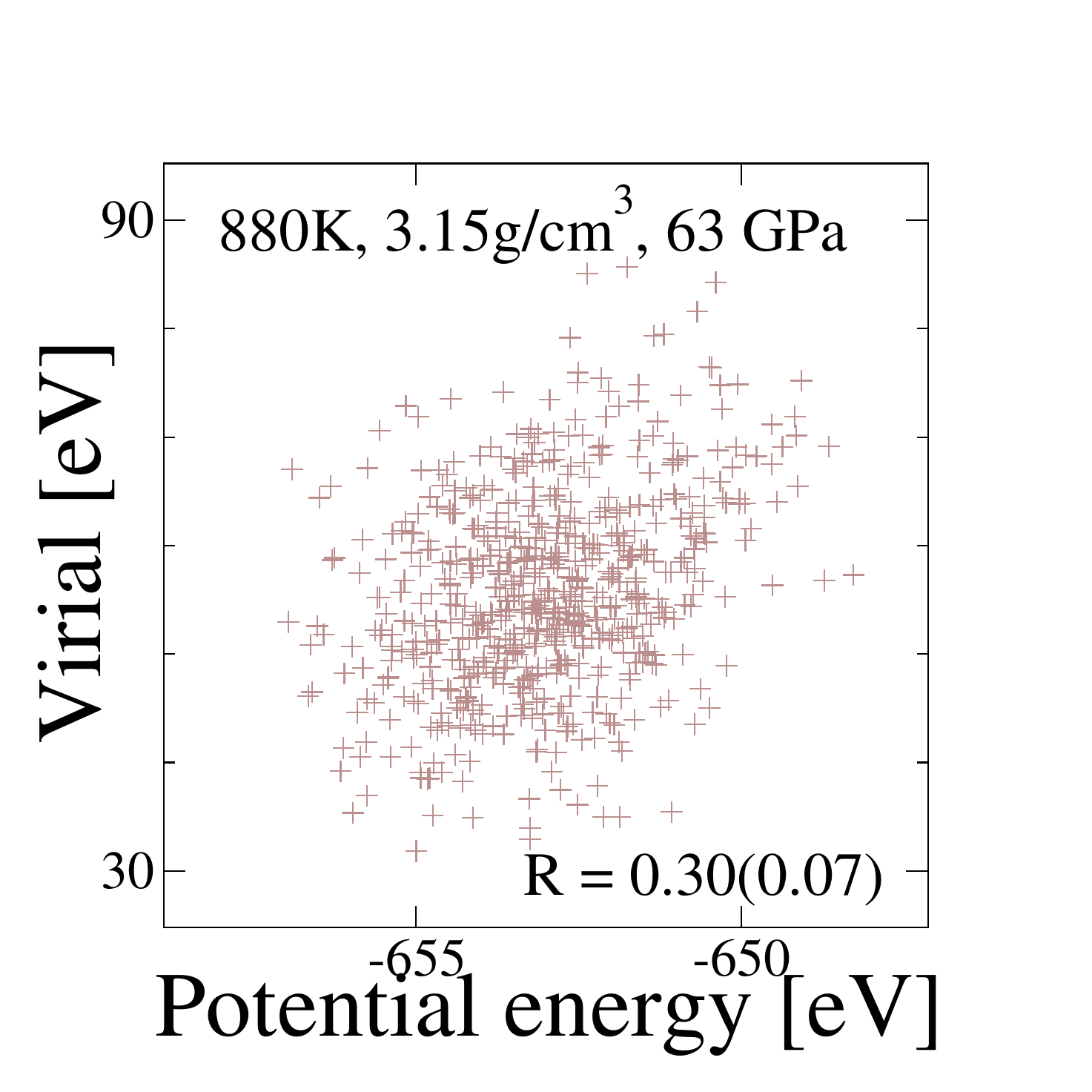}
  \includegraphics[height=0.45\columnwidth]{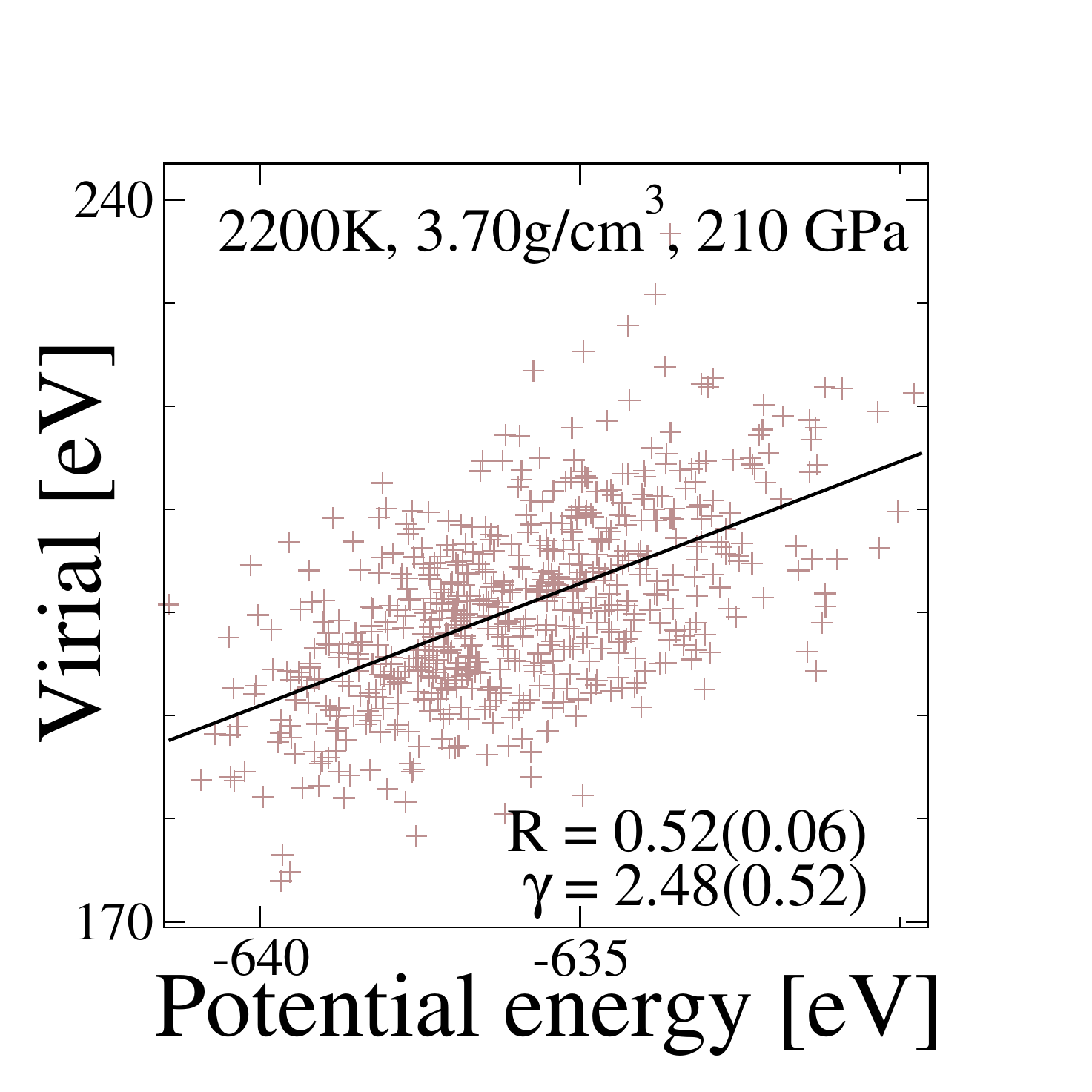}
  \caption{
    High-pressure calculations of phosphorus (P) show that the correlation
    coefficient increases with increasing pressure. 
  }
  \label{fig:PUW}
\end{figure}

\section{Concluding remarks}\label{discussion}

\subsection{The Gr{\"u}neisen equation of state}\label{gruneisen}
It has been known for a long time that at pressures high enough to result in non-negligible compression, solids and liquids generally obey the Gr{\"u}neisen equation of state from 1912 (also referred to as the Mie-Gr{\"u}neisen equation) \cite{bor39,nag11}. This expresses proportionality between pressure and energy $E$ per volume $V$ as follows: 
\be
p=\gamma_{\rm G}(\rho)E/V+C(\rho)
\ee
in which $\gamma_{\rm G}(\rho)$ is the Gr{\"u}neisen parameter (Sec. \ref{subsection:gamma}) and $C(\rho)$ the ``cold pressure'', both of which are functions only of the density. The Gr{\"u}neisen equation has been applied for describing condensed matter in a wide variety of high-pressure situations, ranging from the core of the Earth \cite{poirier} to various forms of explosions \cite{mar02,nag11}. The well-proven Hugoniot shock-adiabatic method is available for determining $\gamma_{\rm G}(\rho)$ experimentally \cite{nag11,landau_fluid,mar80}.
At high pressure in the dense fluid phase not too far from the melting line \cite{bra12} the virial term dominates. For instance, the ideal-gas contribution to the pressure is below five percent throughout the liquid outer core of the Earth. The configuration-space analog of the Gr{\"u}neisen equation is the relation
\be\label{WUgru}
W \cong \gamma(\rho)U+C(\rho)V\,.
\ee
Equation (\ref{WUgru}) follows from the hidden scale invariance \cite{III}. The scaling exponent $\gamma(\rho)$ has a simple relation to $\gamma_{\rm G}(\rho)$ given in Eq.\ (\ref{exponent}) \cite{III}.

\subsection{Empirical melting and freezing rules}
As a consequence of our finding, many empirical melting and freezing rules now find a concise explanation. Specifically, a number of invariants along the melting line of metals (and model systems) have been known for years with no good explanations. These rules follow from hidden scale invariance, because the melting and freezing lines are both isomorphs \cite{IV} and the rules all involve isomorph invariants. A famous melting rule is the Lindemann criterion, according to which a crystal melts when the thermal vibrational atomic displacement is about 10\% of the crystal's interatomic distance \cite{gil56,ubb65,ros69,sti75}. While our work does not imply a universal value of 10\%, it does imply invariance of the Lindemann quantity along the melting line. There are also other empirically well-established freezing rules of invariance, for instance the Hansen-Verlet rule that a liquid crystallizes when the first peak of the structure factor reaches the value 2.85 \cite{han69}, the Andrade equation predicting constant reduced-unit viscosity along the freezing line \cite{and34,kap05}, the Raveche-Mountain-Streett criterion \cite{rav74} of a quasiuniversal ratio between maximum and minimum of the radial distribution function at freezing, Lyapunov-exponent based criteria \cite{mal00}, or the criterion of zero higher-than-second-order liquid configurational entropy at crystallization \cite{sai01}. Connecting the melting and freezing lines is the rule of invariant constant-volume melting entropy \cite{tal80,wallace}.



\acknowledgments 
The authors are indebted to Nick Bailey for illuminating discussions. 
This work was financially supported by the Austrian Science Fund FWF within the SFB ViCoM (F41).
The center for viscous liquid dynamics ``Glass and Time'' is sponsored by the Danish National Research Foundation's grant DNRF61.
U.R.P. was supported by the Villum Foundation's grand VKR-023455.
The Vienna Scientific Cluster (VSC) was used for computations.

\appendix
\section{Estimations of DFT triple points}
\begin{figure}
  \includegraphics[height=0.30\columnwidth]{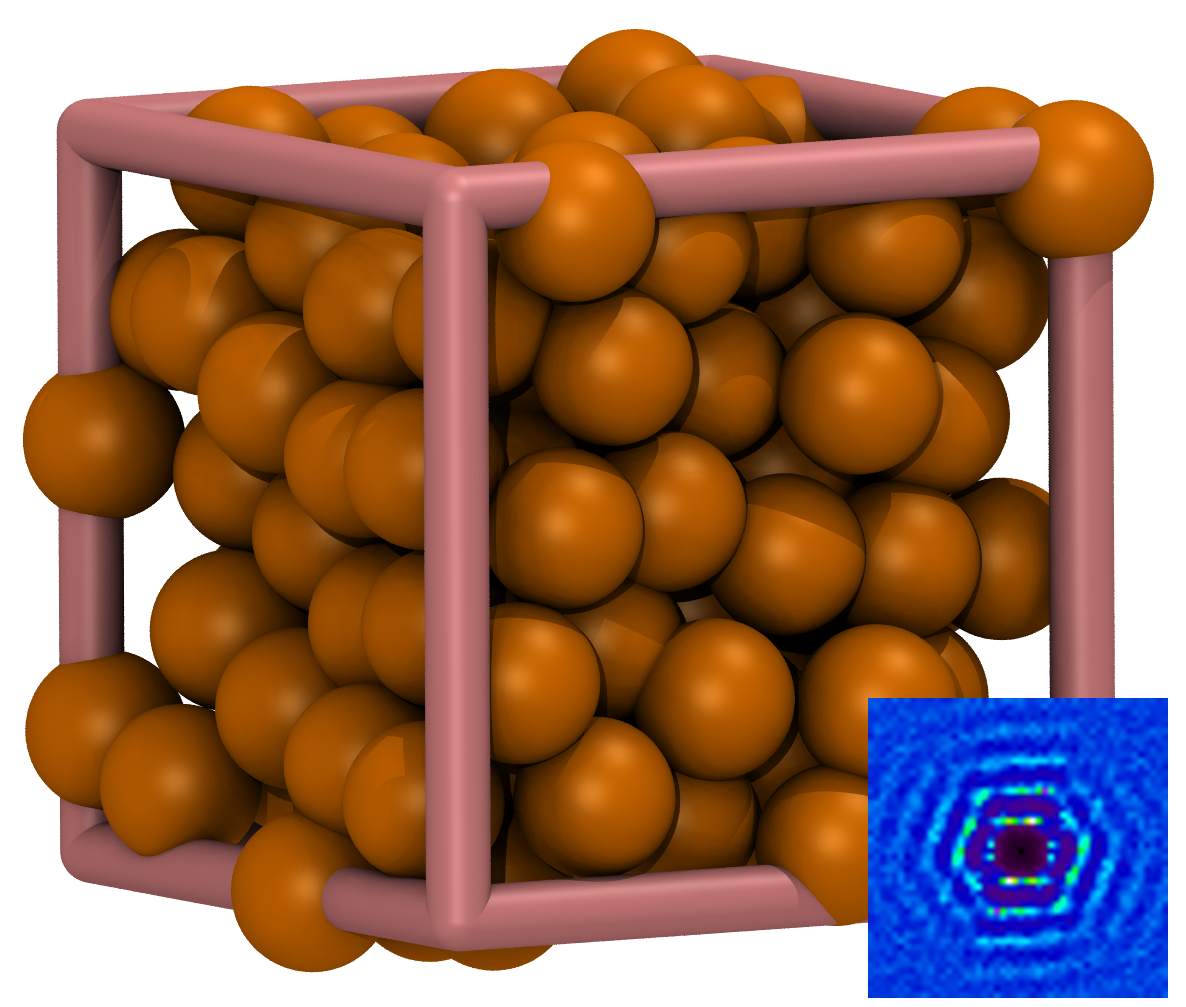}
  \includegraphics[height=0.40\columnwidth]{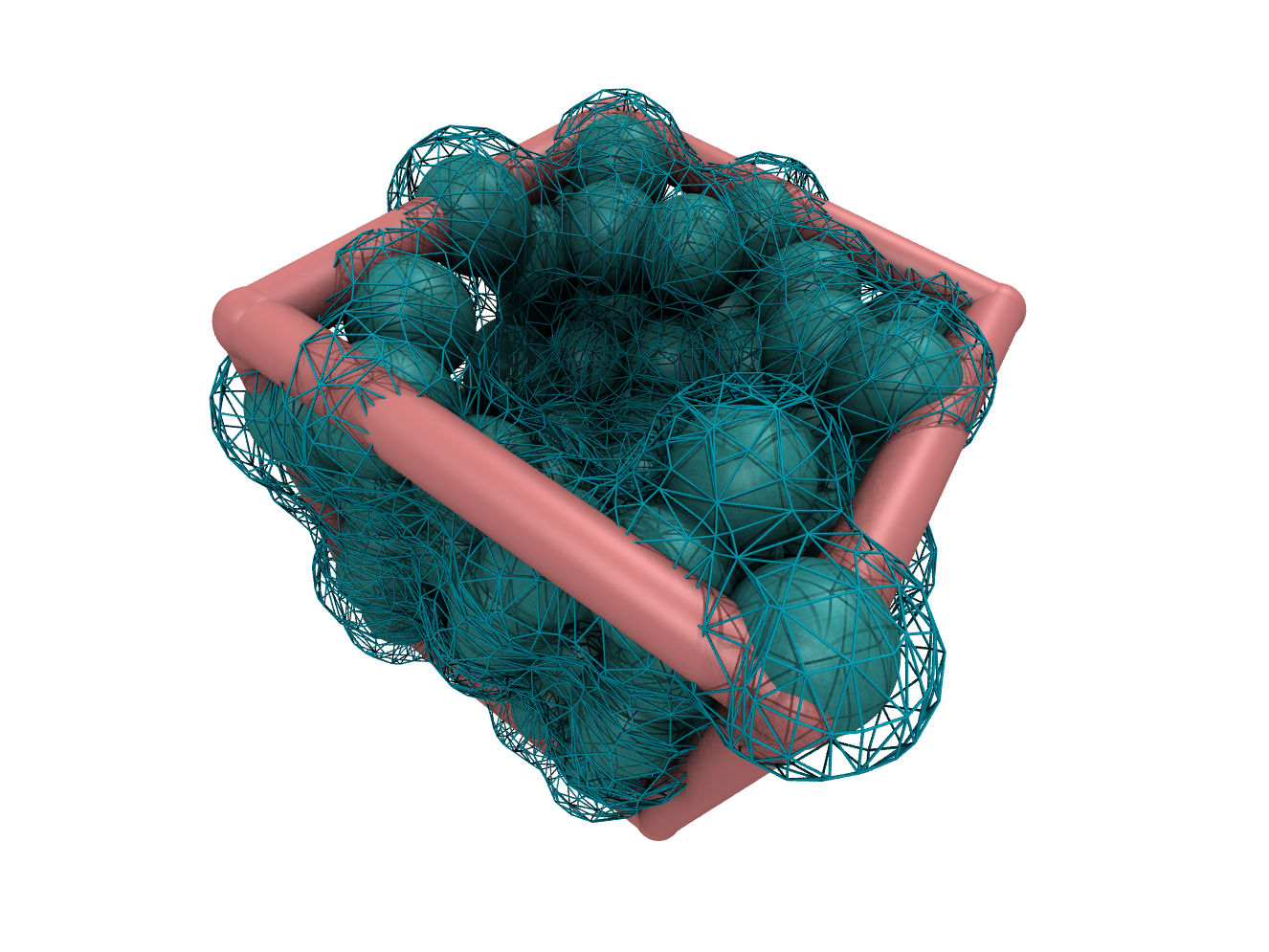}
 \caption{ Panel (a) show an example of a crystallized configuration (Ga). The inset shows the computed diffraction patten. Whenever this happened, computations was redone at higher temperature to avoid crystallization. Panel (b) shows a DFT configuration of Fe at the experimental triple point. DFT overestimates the triple point density by 16\%; thus cavities are formed when iron is simulated at the experimental density.}
  \label{fig:Configurations}
\end{figure}

The main focus of the paper is to determine hidden scale invariances of elements in the 
low-pressure part of the liquid phase.
An unbiased choice is to use state points near the DFT triple points. However, an accurate determination of the triple point is beyond the scope of the present paper,
and also not relevant for the present work. 
The DFT triple point is estimated by performing an $NpT$ computation at a state point close to, or slightly above, the experimental melting temperature and ambient pressure (within an accuracy of 2\,GPa or less, see Table \ref{tab:details}).
The $NpT$ ensemble is realized using  the Parinello-Rahman method
\cite{par81}
with a fictitious mass of 100 atomic units and a thermal coupling time of 0.33\,ps.
The finite cut-off of the plane wave basis set requires the exertion of an
additional pressure to compensate for the missing basis set functions. This
pressure is referred to as Pulay stress and it was estimated in a separate
calculation. 
If the system froze at the experimental temperature, for instance due to
finite size effects, the temperature was increased until the system
stayed liquid throughout the entire trajectory. An example of a crystallized configuration is shown on Fig.\ \ref{fig:Configurations}(a). For some systems, such as Ti,
Ga, Sn and Hg the temperature had to be increased significantly compared to the
experimental melting temperature. However, we only found small variations of
$\gamma$ and $R$ upon increasing the  temperature by a factor 2.
%
The final state points used to compute the $UW$ correlations are
listed in Tab \ref{tab:details}.

For Mn, Fe, and Co the computed melting density is 25\%, 16\%
and 15\% higher than the experimental density. 
Simulations of these liquids at the experimental density using
standard density functional theory led to internal surfaces and 
cavities between the atoms as shown in Fig.\ \ref{fig:Configurations}(b).
The reason for this sizable
deviation of the melting densities from experiment needs to be investigated
in the future. A likely explanations is strong paramagnetic fluctuations at the atomic
sites that would require a treatment beyond standard density functional theory,
for instance using dynamical mean field theory. Such fluctuations
can increase the typical bond length by some 5\%.

\bibliography{abiniipl}

\begin{thebibliography}{100}

\bibitem{bak2002}
P.~Bak, K.~Christensen, L.~Danon, and T.~Scanlon, ``Unified scaling law for
  earthquakes,'' {\em Phys. Rev. Lett.}, vol.~88, p.~178501, Apr 2002.

\bibitem{mandelbrot1968}
B.~B. Mandelbrot and J.~W.~V. Ness, ``Fractional brownian motions, fractional
  noises and applications,'' {\em SIAM Rev.}, vol.~10, no.~4, pp.~422--437,
  1968.

\bibitem{guth1982}
A.~H. Guth and S.-Y. Pi, ``Fluctuations in the new inflationary universe,''
  {\em Phys. Rev. Lett.}, vol.~49, pp.~1110--1113, Oct 1982.

\bibitem{hawking1982}
S.~Hawking, ``The development of irregularities in a single bubble inflationary
  universe,'' {\em Phys. Lett. B}, vol.~115, no.~4, pp.~295 -- 297, 1982.

\bibitem{mandelbrot83}
B.~B. Mandelbrot, {\em The Fractal Geometry of Nature}.
\newblock Freeman, 1983.

\bibitem{pelissetto2002}
A.~Pelissetto and E.~Vicari, ``Critical phenomena and renormalization-group
  theory,'' {\em Phys. Rep.}, vol.~368, no.~6, pp.~549 -- 727, 2002.

\bibitem{ped08}
U.~R. Pedersen, N.~P. Bailey, T.~B. Schr{\o}der, and J.~C. Dyre, ``{Strong
  Pressure-Energy Correlations in van der Waals Liquids},'' {\em Phys. Rev.
  Lett.}, vol.~100, p.~015701, 2008.

\bibitem{I}
N.~P. Bailey, U.~R. Pedersen, N.~Gnan, T.~B. Schr{\o}der, and J.~C. Dyre,
  ``{Pressure-Energy Correlations in Liquids. I. Results from Computer
  Simulations},'' {\em J. Chem. Phys.}, vol.~129, p.~184507, 2008.

\bibitem{II}
N.~P. Bailey, U.~R. Pedersen, N.~Gnan, T.~B. Schr{\o}der, and J.~C. Dyre,
  ``{Pressure-Energy Correlations in Liquids. II. Analysis and Consequences},''
  {\em J. Chem. Phys.}, vol.~129, p.~184508, 2008.

\bibitem{III}
T.~B. Schr{\o}der, N.~P. Bailey, U.~R. Pedersen, N.~Gnan, and J.~C. Dyre,
  ``{Pressure-Energy Correlations in Liquids. III. Statistical Mechanics and
  Thermodynamics of Liquids with Hidden Scale Invariance},'' {\em J. Chem.
  Phys.}, vol.~131, p.~234503, 2009.

\bibitem{IV}
N.~Gnan, T.~B. Schr{\o}der, U.~R. Pedersen, N.~P. Bailey, and J.~C. Dyre,
  ``{Pressure-Energy Correlations in Liquids. IV. “Isomorphs” in Liquid
  Phase Diagrams},'' {\em J. Chem. Phys.}, vol.~131, p.~234504, 2009.

\bibitem{ing12}
T.~S. Ingebrigtsen, T.~B. Schr\o{}der, and J.~C. Dyre, ``What is a simple
  liquid?,'' {\em Phys. Rev. X}, vol.~2, p.~011011, Mar 2012.

\bibitem{dyr14}
J.~C. Dyre, ``{Hidden Scale Invariance in Condensed Matter},'' {\em J. Phys.
  Chem. B}, vol.~118, pp.~10007--10024, 2014.

\bibitem{mal13}
A.~Malins, J.~Eggers, and C.~P. Royall, ``{Investigating Isomorphs with the
  Topological Cluster Classification},'' {\em J. Chem. Phys.}, vol.~139,
  p.~234505, 2013.

\bibitem{pra14}
S.~Prasad and C.~Chakravarty, ``{Onset of Simple Liquid Behaviour in Modified
  Water Models},'' {\em J. Chem. Phys.}, vol.~140, no.~16, p.~164501, 2014.

\bibitem{fle14}
E.~Flenner, H.~Staley, and G.~Szamel, ``{Universal Features of Dynamic
  Heterogeneity in Supercooled Liquids},'' {\em Phys. Rev. Lett.}, vol.~112,
  p.~097801, 2014.

\bibitem{hen14}
A.~Henao, S.~Pothoczki, M.~Canales, E.~Guardia, and L.~Pardo, ``Competing
  structures within the first shell of liquid {${\rm C_2Cl_6}$}: A molecular
  dynamics study,'' {\em J. Mol. Liquids}, vol.~190, pp.~121--125, 2014.

\bibitem{pie14}
S.~Pieprzyk, D.~M. Heyes, and A.~C. Branka, ``Thermodynamic properties and
  entropy scaling law for diffusivity in soft spheres,'' {\em Phys. Rev. E},
  vol.~90, p.~012106, 2014.

\bibitem{hey15}
D.~M. Heyes, D.~Dini, and A.~C. Branka, ``Scaling of lennard-jones liquid
  elastic moduli, viscoelasticity and other properties along fluid-solid
  coexistence,'' {\em Phys. Status Solidi (b)}, pp.~n/a--n/a, 2015.

\bibitem{gun11}
D.~Gundermann, U.~R. Pedersen, T.~Hecksher, N.~P. Bailey, B.~Jakobsen,
  T.~Christensen, N.~B. Olsen, T.~B. Schr{\o}der, D.~Fragiadakis, R.~Casalini,
  C.~M. Roland, J.~C. Dyre, and K.~Niss, ``{Predicting the Density-Scaling
  Exponent of a Glass--Forming Liquid from Prigogine–-Defay Ratio
  Measurements},'' {\em Nature Phys.}, vol.~7, pp.~816--821, 2011.

\bibitem{rol05}
C.~M. Roland, S.~Hensel-Bielowka, M.~Paluch, and R.~Casalini, ``{Supercooled
  Dynamics of Glass-Forming Liquids and Polymers under Hydrostatic Pressure},''
  {\em Rep. Prog. Phys.}, vol.~68, pp.~1405--1478, 2005.

\bibitem{rol10}
C.~M. Roland, ``{Relaxation Phenomena in Vitrifying Polymers and Molecular
  Liquids},'' {\em Macromolecules}, vol.~43, pp.~7875--7890, 2010.

\bibitem{alb14}
D.~E. Albrechtsen, A.~E. Olsen, U.~R. Pedersen, T.~B. Schr{\o}der, and J.~C.
  Dyre, ``Isomorph invariance of the struccture and dynamics of classical
  crystals,'' {\em Phys. Rev. B}, vol.~90, p.~094106, 2014.

\bibitem{waals1873}
J.~D. van~der Waals, {\em On the Continuity of the Gaseous and Liquid States}.
\newblock PhD thesis, Universiteit Leiden, 1873.

\bibitem{bor32}
M.~Born and J.~E. Meyer, ``{Zur Gittertheorie der Ionenkristalle},'' {\em Z.
  Phys.}, vol.~75, pp.~1--18, 1932.

\bibitem{bor39}
M.~Born, ``Thermodynamics of crystals and melting,'' {\em J. Chem. Phys.},
  vol.~7, pp.~591--603, 1939.

\bibitem{wid67}
B.~Widom, ``Intermolecular forces and the nature of the liquid state,'' {\em
  Science}, vol.~157, p.~375, 1967.

\bibitem{wca}
J.~D. Weeks, D.~Chandler, and H.~C. Andersen, ``Role of repulsive forces in
  determining the equilibrium structure of simple liquids,'' {\em J. Chem.
  Phys.}, vol.~54, pp.~5237--5247, 1971.

\bibitem{gub71}
K.~Gubbins, W.~Smitha, M.~Tham, and E.~Tiepel, ``Perturbation theory for the
  radial distribution function,'' {\em Mol. Phys..}, vol.~22, p.~1089, 1971.

\bibitem{bar76}
J.~A. Barker and D.~Henderson, ``What is "liquid"? {Understanding} the states
  of matter,'' {\em Rev. Mod. Phys.}, vol.~48, pp.~587--671, 1976.

\bibitem{and76}
H.~C. Andersen, D.~Chandler, and J.~D. Weeks, ``Role of repulsive and
  attractive forces in liquids: the equlibrium theory of classical fluids,''
  {\em Adv. Chem. Phys.}, vol.~34, p.~105, 1976.

\bibitem{wri81}
C.~C. Wright and P.~T. Cummings, ``A new perturbation theory for potentials
  with a soft core. application to liquid sodium,'' {\em Chem. Phys. Lett.},
  vol.~83, pp.~120--124, 1981.

\bibitem{hoo71}
W.~G. Hoover, S.~G. Gray, and K.~W. Johnson, ``Thermodynamic properties of the
  fluid and solid phases for inverse power potentials,'' {\em J. Chem. Phys.},
  vol.~55, pp.~1128--1136, 1971.

\bibitem{hoo72}
W.~G. Hoover, D.~A. Young, and R.~Grover, ``Statistical mechanics of phase
  diagrams. i. inverse power potentials and the close-packed to body-packed
  cubic transition,'' {\em J. Chem. Phys.}, vol.~56, pp.~2207--2210, 1972.

\bibitem{hiw74}
Y.~Hiwatari, H.~Matsuda, T.~Ogawa, N.~Ogita, and A.~Ueda, ``Molecular dynamics
  studies on the soft-core model,'' {\em Prog. Theor. Phys.}, vol.~52,
  pp.~1105--1123, 1974.

\bibitem{sti75}
S.~M. Stishov, ``{The Thermodynamics of Melting of Simple Substances},'' {\em
  Sov. Phys. Usp.}, vol.~17, no.~5, pp.~625--643, 1975.

\bibitem{ros76}
Y.~Rosenfeld, ``Universality of melting and freezing indicators and additivity
  of melting curves,'' {\em Mol. Phys.}, vol.~32, pp.~963--977, 1976.

\bibitem{you77}
D.~A. Young, ``A soft-sphere model for liquid metals,'' tech. rep., Lawrence
  Livermore Laboratorie, 1977.

\bibitem{ros83}
Y.~Rosenfeld, ``Variational soft-sphere perturbation theory and conditions for
  a gruneisen equation of state for dense fluids,'' {\em Phys. Rev. A},
  vol.~28, pp.~3063--3069, 1983.

\bibitem{you91}
D.~A. Young, {\em Phase Diagram of the Elements}.
\newblock University of California Press, 1991.

\bibitem{pre05}
S.~Prestipino, F.~Saija, and P.~V. Giaquinta, ``Phase diagram of softly
  repulsive systems: The {Gaussian} and inverse-power-law potentials,'' {\em J.
  Chem. Phys.}, vol.~123, p.~144110, 2005.

\bibitem{bra06}
A.~C. Branka and D.~M. Heyes, ``Thermodynamic properties of inverse power
  fluids,'' {\em Phys. Rev. E}, vol.~74, p.~031202, 2006.

\bibitem{hey07}
D.~M. Heyes and A.~C. Branka, ``Physical properties of soft repulsive particle
  fluids,'' {\em Phys. Chem. Chem. Phys.}, vol.~9, pp.~5570--5575, 2007.

\bibitem{hey08}
D.~M. Heyes and A.~C. Branka, ``Self-diffusion coefficients and shear viscosity
  of inverse power fluids: from hard- to soft-spheres,'' {\em Phys. Chem. Chem.
  Phys.}, vol.~10, pp.~4036--4044, 2008.

\bibitem{ped10}
U.~R. Pedersen, T.~B. Schr{\o}der, and J.~C. Dyre, ``Repulsive reference
  potential reproducing the dynamics of a liquid with attractions,'' {\em Phys.
  Rev. Lett.}, vol.~105, p.~157801, 2010.

\bibitem{bra11}
A.~C. Branka and D.~M. Heyes, ``Pair correlation function of soft-sphere
  fluids,'' {\em J. Chem. Phys.}, vol.~134, p.~064115, 2011.

\bibitem{khr11a}
S.~A. Khrapak, M.~Chaudhuri, and G.~E. Morfill, ``{Communication: Universality
  of the Melting Curves for a Wide Range of Interaction Potentials},'' {\em J.
  Chem. Phys.}, vol.~134, p.~241101, 2011.

\bibitem{tra14}
A.~Travesset, ``Phase diagram of power law and lennard-jones systems: Crystal
  phases,'' {\em J. Chem Phys.}, 2014.

\bibitem{hafner86}
J.~Hafner, ``Electronic aspects of the structure and of the glass-forming
  abilities of metallic alloys,'' in {\em Amorphous metals and simiconductors}
  (P.~Hansen and R.~I. Jaffee, eds.), vol.~3 of {\em Acta-scripta metallurgica
  proceedings}, p.~151, Pergamon press, 1986.

\bibitem{koh99}
W.~Kohn, ``Nobel lecture: Electronic structure of matter - wave functions and
  density functionals,'' {\em Rev. Mod. Phys.}, vol.~71, pp.~1253--1266, Oct
  1999.

\bibitem{pop99}
J.~A. Pople, ``Nobel lecture: Quantum chemical models,'' {\em Rev. Mod. Phys.},
  vol.~71, pp.~1267--1274, Oct 1999.

\bibitem{coh12}
A.~J. Cohen, P.~Mori-Sánchez, and W.~Yang, ``Challenges for density functional
  theory,'' {\em Chem. Rev.}, vol.~112, no.~1, pp.~289--320, 2012.

\bibitem{bur12}
K.~Burke, ``Perspective on density functional theory,'' {\em J. Chem Phys.},
  vol.~136, p.~150901, 2012.

\bibitem{alb02}
C.~Alba-Simionesco, D.~Kivelson, and G.~Tarjus, ``{Temperature, Density, and
  Pressure Dependence of Relaxation Times in Supercooled Liquids},'' {\em J.
  Chem. Phys.}, vol.~116, pp.~5033--5038, 2002.

\bibitem{alb06}
C.~Alba-Simionesco and G.~Tarjus, ``Temperature versus density effects in
  glassforming liquids and polymers: A scaling hypothesis and its
  consequences,'' {\em J. Non-Cryst. Solids}, vol.~352, p.~4888, 2006.

\bibitem{flo11}
G.~Floudas, M.~Paluch, A.~Grzybowski, and K.~Ngai, {\em {Molecular Dynamics of
  Glass-Forming Systems: Effects of Pressure}}.
\newblock Springer, Berlin, 2011.

\bibitem{mau14}
P.~Mausbach and H.-O. May, ``{Direct Molecular Simulation of the Gr{\"u}neisen
  Parameter and Density Scaling Exponent in Fluid Systems},'' {\em Fluid Phase
  Equilibria}, vol.~366, no.~0, pp.~108--116, 2014.

\bibitem{han13}
J.-P. Hansen and I.~R. McDonald, {\em {Theory of Simple Liquids: With
  Applications to Soft Matter}}.
\newblock Academic, New York, fourth~ed., 2013.

\bibitem{tildesley}
M.~P. Allen and D.~J. Tildesley, {\em {Computer Simulation of Liquids}}.
\newblock Oxford Science Publications, 1987.

\bibitem{PhysRevLett.114.195901}
A.~Glensk, B.~Grabowski, T.~Hickel, and J.~Neugebauer, ``Understanding
  anharmonicity in fcc materials: From its origin to \textit{ab initio}
  strategies beyond the quasiharmonic approximation,'' {\em Phys. Rev. Lett.},
  vol.~114, p.~195901, May 2015.

\bibitem{PhysRevLett.114.166101}
B.~Jiang and H.~Guo, ``Dynamics of water dissociative chemisorption on ni(111):
  Effects of impact sites and incident angles,'' {\em Phys. Rev. Lett.},
  vol.~114, p.~166101, Apr 2015.

\bibitem{PhysRevB.88.094101}
U.~R. Pedersen, F.~Hummel, G.~Kresse, G.~Kahl, and C.~Dellago, ``{Computing
  Gibbs free energy differences by interface pinning},'' {\em Phys. Rev. B},
  vol.~88, p.~094101, Sep 2013.

\bibitem{PhysRevB.54.11169}
G.~Kresse and J.~Furthm\"uller, ``Efficient iterative schemes for ab initio
  total-energy calculations using a plane-wave basis set,'' {\em Phys. Rev. B},
  vol.~54, pp.~11169--11186, Oct 1996.

\bibitem{PhysRevB.50.17953}
P.~E. Bl\"ochl, ``Projector augmented-wave method,'' {\em Phys. Rev. B},
  vol.~50, pp.~17953--17979, Dec 1994.

\bibitem{PhysRevLett.100.136406}
J.~P. Perdew, A.~Ruzsinszky, G.~I. Csonka, O.~A. Vydrov, G.~E. Scuseria, L.~A.
  Constantin, X.~Zhou, and K.~Burke, ``Restoring the density-gradient expansion
  for exchange in solids and surfaces,'' {\em Phys. Rev. Lett.}, vol.~100,
  p.~136406, Apr 2008.

\bibitem{crc12}
W.~M. Haynes, ed., {\em CRC Handbook of Chemistry and Physics}.
\newblock Taylor \& Francis, 93rd edition~ed., 2012.

\bibitem{ass12}
M.~J. Assael, I.~J. Armyra, J.~Brillo, S.~V. Stankus, J.~Wu, and W.~A. Wakeham,
  ``Reference data for the density and viscosity of liquid cadmium, cobalt,
  gallium, indium, mercury, silicon, thallium, and zinc,'' {\em J. Phys. Chem.
  Ref. Data}, vol.~41, p.~033101, Sept. 2012.

\bibitem{sav05}
A.~I. Savvatimskiy, ``Measurements of the melting point of graphite and the
  properties of liquid carbon (a review for 1963–2003),'' {\em Carbon},
  vol.~43, no.~6, pp.~1115--1142, 2005.

\bibitem{haa76}
D.~M. Haaland, ``Graphite-liquid-vapor triple point pressure and the density of
  liquid carbon,'' {\em Carbon}, vol.~14, no.~6, pp.~357--361, 1976.

\bibitem{tru00}
T.~M. Truskett, S.~Torquato, and P.~G. Debenedetti, ``{Towards a Quantification
  of Disorder in Materials: Distinguishing Equilibrium and Glassy Sphere
  Packings},'' {\em Phys. Rev. E}, vol.~62, pp.~993--1001, Jul 2000.

\bibitem{sin07}
R.~N. Singh, S.~Arafin, and A.~K. George, ``Temperature-dependent
  thermo-elastic properties of s-, p- and d-block liquid metals,'' {\em Physica
  B}, 2007.

\bibitem{her99}
J.~N. Herrera, P.~T. Cummings, and H.~Ruiz-Estrada, ``Static structure factor
  for simple liquid metals,'' {\em Molecular Physics}, vol.~96, no.~5,
  pp.~835--847, 1999.

\bibitem{tah06}
S.~Tahara, H.~Fujii, Y.~Yokota, Y.~Kawakita, S.~Kohara, and S.~Takeda,
  ``Structure and electron-ion correlation in liquid {Mg},'' {\em Phys. B},
  vol.~385, no.~0, pp.~219--221, 2006.
\newblock Proceedings of the Eighth International Conference on Neutron
  Scattering.

\bibitem{sen09}
S.~Seng{\"u}l, D.~J. González, and L.~E. González, ``Structural and dynamical
  properties of liquid mg. an orbital-free molecular dynamics study,'' {\em J.
  Phys. Condens. Matter}, vol.~21, no.~11, p.~115106, 2009.

\bibitem{mcl82}
I.~L. McLaughlin and W.~H. Young, ``Calculation of the small and large angle
  structure factors of some simple liquid metals,'' {\em J. Phys}, vol.~12,
  p.~245, 1982.

\bibitem{lai92}
B.~B. Laird and A.~D.~J. Haymet, ``Phase diagram for the inverse sixth power
  potential system from molecular dynamics computer simulation,'' {\em Mol.
  Phys.}, vol.~75, no.~1, p.~71, 1992.

\bibitem{agr95}
R.~Agrawal and D.~A. Kofke, ``Solid-fluid coexistence for inverse-power
  potentials,'' {\em Phys. Rev. Lett.}, vol.~74, p.~122, 1995.

\bibitem{ton04}
E.~Y. Tonkov and E.~G. Ponyatovsky, {\em Phase Transformations of Elements
  Under High Pressure}.
\newblock CRC Press, 2004.

\bibitem{gri12}
G.~Grimvall, B.~Magyari-Koepe, V.~Ozolins, and K.~A. Persson, ``Lattice
  instabilities in metallic elements,'' {\em Rev. Mod. Phys.}, vol.~84,
  pp.~945--986, 2012.

\bibitem{ped13}
U.~R. Pedersen, ``Direct calculation of the solid-liquid {Gibbs} free energy
  difference in a single equilibrium simulation,'' {\em J. Chem. Phys.},
  vol.~139, p.~104102, 2013.

\bibitem{V}
T.~B. Schr{\o}der, N.~Gnan, U.~R. Pedersen, N.~P. Bailey, and J.~C. Dyre,
  ``{Pressure-Energy Correlations in Liquids. V. Isomorphs in Generalized
  Lennard-Jones Systems},'' {\em J. Chem. Phys.}, vol.~134, p.~164505, 2011.

\bibitem{poirier}
J.-P. Poirier, {\em Introduction to the Physics of the Earth's Interior}.
\newblock Cambridge Univer-sity Press, 2000.

\bibitem{PhysRevB.87.094102}
J.~Bouchet, S.~Mazevet, G.~Morard, F.~Guyot, and R.~Musella, ``{{\it Ab initio}
  equation of state of iron up to 1500 GPa},'' {\em Phys. Rev. B}, vol.~87,
  p.~094102, Mar 2013.

\bibitem{nag11}
K.~Nagayama, {\em {Introduction to the Gr{\"u}neisen Equation of State and
  Shock Thermodynamics}}.
\newblock Amazon Digital Services, Inc., 2011.

\bibitem{mar02}
N.~H. March and M.~P. Tosi, {\em Introduction to the Liquid State}.
\newblock World Scientific, Singapore, 2002.

\bibitem{landau_fluid}
L.~D. Landau and E.~M. Lifshitz, {\em Fluid Mechanics}.
\newblock Pergamon, Oxford, 1959.

\bibitem{mar80}
S.~P. Marsh, ed., {\em LASL shock Hugoniot data}.
\newblock University of California Press, 1980.

\bibitem{bra12}
V.~V. Brazhkin, Y.~D. Fomin, A.~G. Lyapin, V.~N. Ryzhov, and K.~Trachenko,
  ``Two liquid states of matter: A dynamic line on a phase diagram,'' {\em
  Phys. Rev. E}, vol.~85, p.~031203, Mar 2012.

\bibitem{gil56}
J.~J. Gilvarry, ``The {Lindemann} and {G}r\"uneisen laws,'' {\em Phys. Rev.},
  vol.~102, pp.~308--316, Apr 1956.

\bibitem{ubb65}
A.~R. Ubbelohde, {\em {Melting and Crystal Structure}}.
\newblock Clarendon, Oxford, 1965.

\bibitem{ros69}
M.~Ross, ``{Generalized Lindemann melting law},'' {\em Phys. Rev.}, vol.~184,
  pp.~233--242, 1969.

\bibitem{han69}
J.-P. Hansen and L.~Verlet, ``Phase transitions of the {Lennard-Jones}
  system,'' {\em Phys. Rev.}, vol.~184, pp.~151--161, 1969.

\bibitem{and34}
{E. N. C. Andrade}, ``{A theory of the viscosity of liquids - Part I},'' {\em
  Phil. Mag.}, vol.~17, pp.~497--511, 1934.

\bibitem{kap05}
G.~Kaptay, ``A unified equation for the viscosity of pure liquid metals,'' {\em
  Z. Metallkd.}, vol.~96, pp.~24--31, 2005.

\bibitem{rav74}
H.~J. Raveche, R.~D. Mountain, and W.~B. Streett, ``Freezing and melting
  properties of the {Lennard-Jones} system,'' {\em J. Chem. Phys.}, vol.~61,
  pp.~1970--1984, 1974.

\bibitem{mal00}
G.~Malescio, P.~V. Giaquinta, and Y.~Rosenfeld, ``Structural stability of
  simple classical fluids: {Universal} properties of the {Lyapunov-exponent}
  measure,'' {\em Phys. Rev. E}, vol.~61, pp.~4090--4094, 2000.

\bibitem{sai01}
F.~Saija, S.~Prestipino, and P.~V. Giaquinta, ``{Scaling of local density
  correlations in a fluid close to freezing},'' {\em J. Chem. Phys.},
  vol.~{115}, pp.~7586--7591, {2001}.

\bibitem{tal80}
J.~L. Tallon, ``The entropy change on melting of simple substances,'' {\em
  Phys. Lett. A}, vol.~76, pp.~139--142, 1980.

\bibitem{wallace}
D.~C. Wallace, {\em Statistical Physics of Crystals and Liquids}.
\newblock World Scientific, Singapore, 2002.

\bibitem{par81}
M.~Parrinello and A.~Rahman, ``Polymorphic transitions in single crystals: {A}
  new molecular dynamics method,'' {\em J. Appl. Phys.}, vol.~52, no.~12,
  pp.~7182--7190, 1981.

\end{thebibliography}


\end{document}